\documentclass[aip,reprint,author-year,nofootinbib]{revtex4-1}

\usepackage{graphicx}
\usepackage{amsmath,amssymb}             
\usepackage{stmaryrd}					
\usepackage{color}
\usepackage[dvipsnames]{xcolor}
\usepackage{color,hyperref}
\definecolor{darkblue}{rgb}{0.0,0.0,0.4}
\definecolor{green}{rgb}{0.0,0.5,0.0}
\hypersetup{colorlinks,breaklinks,
            linkcolor=darkblue,urlcolor=darkblue,
            anchorcolor=darkblue,citecolor=RoyalBlue}

\usepackage{gensymb}
\usepackage[mathscr]{euscript}
\usepackage{upgreek}
\DeclareFontFamily{OT1}{pzc}{}
\DeclareFontShape{OT1}{pzc}{m}{it}{<-> s * [1.10] pzcmi7t}{}
\DeclareMathAlphabet{\mathpzc}{OT1}{pzc}{m}{it}
\usepackage{stmaryrd}
\usepackage{tikz}
\usetikzlibrary{shapes,arrows,shadows}

\usepackage{chngcntr}
\usepackage{ulem}
\usepackage{makecell}

\def\apj{Astrophys. J.}

\def\mnras{Mon. Not. R. Astron. Soc.}

\def\prd{Phys. Rev. D}

\def\nat{Nature}

\def\aap{Astron. \& Astrophys.}

\def\jcap{Journal of Cosmology and Astroparticle Physics}

\def\nar{New Astronomy Reviews}
\def\pasp{Publications of the Astronomical Society of the Pacific}

\newcommand{\p}{\mathpzc{P}}
\newcommand{\selfi}{\textsc{selfi}}
\newcommand{\pyselfi}{py\textsc{selfi}}

\begin{document}


\title{
Primordial power spectrum and cosmology from black-box galaxy surveys
}



\author{Florent Leclercq}
\email{florent.leclercq@polytechnique.org}
\homepage{http://www.florent-leclercq.eu/}
\affiliation{Imperial Centre for Inference and Cosmology (ICIC) \& Astrophysics Group, Imperial College London, Blackett Laboratory, Prince Consort Road, London SW7 2AZ, United Kingdom}

\author{Wolfgang Enzi}
\email{enzi@mpa-garching.mpg.de}
\affiliation{Max-Planck Institute for Astrophysics, Karl-Schwarzschild Strasse 1, D-85748 Garching, Germany}

\author{Jens Jasche}
\email{jens.jasche@fysik.su.se}
\affiliation{The Oskar Klein Centre, Department of Physics, Stockholm University, Albanova University Center, SE 106 91 Stockholm, Sweden}
\affiliation{Excellence Cluster Universe, Technische Universit\"at M\"unchen, Boltzmannstrasse 2, D-85748 Garching, Germany}

\author{Alan Heavens}
\email{a.heavens@imperial.ac.uk}
\affiliation{Imperial Centre for Inference and Cosmology (ICIC) \& Astrophysics Group, Imperial College London, Blackett Laboratory, Prince Consort Road, London SW7 2AZ, United Kingdom}


\date{\today}

\begin{abstract}
\noindent We propose a new, likelihood-free approach to inferring the primordial matter power spectrum and cosmological parameters from arbitrarily complex forward models of galaxy surveys where all relevant statistics can be determined from numerical simulations, i.e. \textit{black-boxes}. Our approach, which we call simulator expansion for likelihood-free inference (\selfi), builds upon approximate Bayesian computation using a novel effective likelihood, and upon the linearisation of black-box models around an expansion point. Consequently, we obtain simple ``filter equations'' for an effective posterior of the primordial power spectrum, and a straightforward scheme for cosmological parameter inference. We demonstrate that the workload is computationally tractable, fixed \textit{a priori}, and perfectly parallel. As a proof of concept, we apply our framework to a realistic synthetic galaxy survey, with a data model accounting for physical structure formation and incomplete and noisy galaxy observations. In doing so, we show that the use of non-linear numerical models allows the galaxy power spectrum to be safely fitted up to at least $k_\mathrm{max} = 0.5$~$h$/Mpc, outperforming state-of-the-art backward-modelling techniques by a factor of $\sim 5$ in the number of modes used. The result is an unbiased inference of the primordial matter power spectrum across the entire range of scales considered, including a high-fidelity reconstruction of baryon acoustic oscillations. It translates into an unbiased and robust inference of cosmological parameters. Our results pave the path towards easy applications of likelihood-free simulation-based inference in cosmology. We have made our code {\pyselfi} and our data products publicly available at \href{http://pyselfi.florent-leclercq.eu}{http://pyselfi.florent-leclercq.eu}.
\end{abstract}


\maketitle



\section{Introduction}
\label{sec:Introduction}

The cosmic large-scale structure constitutes one of the major sources of information for modern cosmology. According to the current paradigm, all observable structures originate from tiny primordial fluctuations, which evolved via gravitational amplification into the presently-observed cosmic web \citep[see e.g.][]{Peebles1980,Peacock1999}. The statistics of the initial density field are measured to be extremely close to Gaussian-distributed \citep[see e.g.][]{Planck2018I}. As a consequence, the primordial matter power spectrum is a very powerful cosmological probe: it is a sufficient statistical summary under the assumption that fluctuations are Gaussian, and -- even if this assumption is violated -- it remains close to capturing all of the information for all models allowed by observations. A particularly important cosmological signature, imprinted on the primordial matter power spectrum at the time of recombination, is baryon acoustic oscillations (BAOs). It constitutes a fixed comoving length scale (a ``standard ruler''), which, when measured at different cosmic times, gives information on the expansion history of the Universe, including the late-time era of accelerated expansion. BAOs are thus one of the main probes to determine the equation of state of a possible dark energy component \citep[see e.g.][]{Eisenstein2005a,Albrecht2006,Percival2007}. A large variety of early-Universe models exhibit a deterministic relation between physical parameters of interest and the primordial matter power spectrum. The latter is therefore an interesting intermediate product for cosmological analyses, allowing parameter inference and model selection to be performed \textit{a posteriori} without (or with minimal) loss of information. It can be seen as a largely agnostic and model-independent parametrisation of cosmological theories, which relies only on weak assumptions (isotropy and gaussianity).

For a long time, measuring the cosmological matter power spectrum has been one of the main goals of galaxy survey data analysis. However, inferring its shape accurately is a challenging task. Various systematic effects such as redshift uncertainties, complex survey geometries, selection effects, missing observations and foreground contamination can greatly hinder the measurement and analysis \citep[see e.g.][]{Ross2012,Jasche2017}. This problem is particularly important for the next generation of optical surveys, such as provided by ESA's Euclid space mission or the Large Synoptic Survey Telescope (LSST), which are expected to be dominated by systematic rather than statistical uncertainty \citep[see e.g.][]{Laureijs2011,LSSTScienceCollaboration2012}. Even if systematic effects arising from the survey strategy were fully understood and controlled, many theoretical challenges would still be present: galaxy biasing, anisotropic clustering in redshift space, and non-linear structure growth at late times, which reduce the detectability of cosmological signatures such as BAOs \citep[see e.g.][]{Meiksin1999,Eisenstein2007a}. Because of the limited reliability of data models, fits of the galaxy power spectrum focus on linear and mildly non-linear scales, using typically a largest wavenumber of $k_\mathrm{max}=0.3$ $h$/Mpc \citep[e.g.][]{Ross2015}. However, the number of modes used in the analysis scales as $k_\mathrm{max}^3$, meaning that any improvement of data models at small scales (such as what can be achieved via numerical simulations instead of perturbation theory) gives access to much more cosmological information.

As a response to theoretical and observational challenges, many approaches to measure the power spectrum have been proposed. They can be divided into two broad categories: backward-modelling approaches (often associated with frequentist statistics, counting the frequencies of measurements in mock catalogues, and associated covariance matrices) and likelihood-based forward-modelling approaches (often associated with Bayesian statistics). Backward-modelling approaches suggest to directly account and correct for relevant effects in observational data as a pre-processing step. For example, the BAO ``reconstruction'' technique removes redshift-space distortions and corrects the density field from bulk motions using the inverse Zel'dovich approximation \citep[e.g.][]{Eisenstein2007,Padmanabhan2012,Doumler2013,Burden2015,White2015}. The end product is an estimator for the primordial matter power spectrum, which can be close to optimal if all relevant effects are modelled \citep{Smith2015,Seljak2017}. As backward-modelling approaches require substantial expert knowledge input, results are often largely model-dependent and with difficult propagation of uncertainties. Importantly, these approaches often rely on fiducial values for the parameters that are the target of the analysis. More recently, several thorough forward-modelling Bayesian approaches have been proposed to jointly infer the three-dimensional matter density field and its power spectrum from galaxy observations, while properly accounting for uncertainties and systematics \citep{Jasche2010b,Jasche2013BIAS,Jasche2015,Granett2015,Jasche2017}. Considerable effort is also put into reconstructing the primordial density field from present observations \citep{Jasche2013BORG,Wang2013,Wang2014a,Jasche2015BORGSDSS,Lavaux2016BORG2MPP,Jasche2019BORGPM,Bos2019}. These methods are likelihood-based, meaning that they solve the exact inference problem by sampling from the target distribution via sophisticated Markov Chain Monte Carlo (MCMC) methods (Gibbs sampling and/or Hamiltonian Monte Carlo). In order to make the likelihood tractable, they have to involve approximations of the data model.

This paper has similar scientific aims but follows a different spirit: the presented method uses likelihood-free forward-modelling. It introduces one variant of approximate Bayesian computation (ABC) and treats the data model as a \textit{black-box} simulator, i.e.  performs inference without necessity to incorporate any knowledge of the data-generating processes into the analysis. This feature renders ABC ideal to infer the primordial matter power spectrum with arbitrarily complex models of galaxy surveys, including a physical treatment of structure formation and the details of observational processes, which cannot be trivially accounted for in likelihood-based statistical approaches. A popular ABC algorithm is likelihood-free rejection sampling, often coupled to Population Monte Carlo \citep[e.g.][]{Ishida2015,Akeret2015,Jennings2017}. More sophisticated approaches known as \textsc{delfi} \citep{Alsing2018} and \textsc{bolfi} \citep{Leclercq2018BOLFI} have also been recently introduced in cosmology. The ABC approach introduced in this work differs from all of the above in two aspects: (\textit{i}) it allows the treatment of a much larger number of parameters (one hundred in this work), which correspond to primordial power spectrum amplitudes at different wavenumbers. To do so, (\textit{ii}) it assumes the availability of a reasonably good guess of the target parameters, based on previous observations. This situation is fairly typical in cosmology, where the allowed space for parameters is already strongly constrained by previous experiments such as the Planck satellite \citep{PlanckCollaboration2015,Planck2018VI}. Under these assumptions, we derive an effective likelihood for the problem. We use an expansion point in parameter space and linearise the black-box around it. The use of finite differencing to compute the gradient of the black-box makes evaluations of the effective likelihood computationally feasible. When further assuming that the prior is Gaussian, we find that the effective posterior distribution for the primordial power spectrum is a Gaussian with mean and covariance matrix given by two simple ``filter equations'' (equations \eqref{eq:filter_mean} and \eqref{eq:filter_var}), which constitute the main result of this work. Finally, we show how to infer parameters of specific cosmological models using the linearised black-box. We propose to call the algorithm ``simulator expansion for likelihood-free inference'' (\selfi).

In order to illustrate the performance of the method, we apply it to a black-box which emulates realistic cosmological data. This black-box is built using \textsc{Simbelmyn\"e} \citep{Leclercq2015ST}, a hierarchical probabilistic simulator to generate synthetic galaxy survey data.\footnote{\textsc{Simbelmyn\"e} is publicly available at \href{https://bitbucket.org/florent-leclercq/simbelmyne/}{https://bitbucket.org/florent-leclercq/simbelmyne}.} The data model involves a full cosmological $N$-body simulation (performed using our implementation of \textsc{cola}, \citealp{Tassev2013}) to evolve the three-dimensional initial density field. It includes a treatment of galaxy bias, redshift-space distortions, survey geometry, selection effects, and instrumental noise. The statistical summary chosen is the estimated power spectrum of the galaxy number count field, as is standard in large-scale structure data analysis, but can be readily extended to include more information. As a result, the inferred primordial matter power spectrum is unbiased across the entire range of Fourier modes considered, and includes in particular BAOs, which were not included in the expansion point. Our analysis demonstrates that by using a fully numerical data model in conjunction with our statistical approach, one can safely fit the galaxy power spectrum even far in the non-linear regime, up to at least $k_\mathrm{max} \approx 0.5$ $h$/Mpc, which provides a factor of $\sim 5$ increase in the number of modes used, with respect to state-of-the-art backward-modelling techniques. We stress that any possible refinement of the data model used in this work does not change the statistical method, and therefore does not affect the validity of the previous statement.

This paper is organised as follows. In section \ref{sec:Method}, we discuss the statistical method and derive the equations for black-box simulation-based inference of the primordial matter power spectrum and cosmological parameters. In section \ref{sec:Data model}, we describe the data-generating model used to test our method. The results obtained by combining the two are discussed in \ref{sec:Results}. We discuss the application of our method and prospects for cosmological data analysis, and provide our conclusions in section \ref{sec:Discussion and conclusion}. Details of the statistical derivations are given in the appendices.

\section{Method}
\label{sec:Method}

This section describes our method for simulation-based inference of the primordial matter power spectrum and cosmological parameters from black-box galaxy surveys. In section \ref{ssec:Design of an effective likelihood for black-box models}, we design an effective likelihood for black-box models. In section \ref{ssec:Linearisation of black-box models}, we exploit previous knowledge, as could have been obtained by earlier cosmological probes, in order to linearise the black-box. We discuss the parametrisation of the primordial matter power spectrum and its prior distribution in section \ref{ssec:The power spectrum prior distribution}. The equations for the effective posterior distribution are given in section \ref{ssec:The power spectrum effective posterior distribution}. In section \ref{ssec:Optimisation of the prior hyperparameters}, we describe how to optimally choose the hyperparameters appearing in the prior. The inference of cosmological parameters from the linearised black-box is discussed in section \ref{ssec:From the power spectrum to cosmological parameters}.

\subsection{Design of an effective likelihood for black-box models}
\label{ssec:Design of an effective likelihood for black-box models}

\begin{table*}
\centering
  \begin{tabular}{c c c c c c}
  \hline 
  Symbol & Meaning & Interpretation\\\hline
  $\boldsymbol{\uptheta} \in \mathbb{R}^S$ & Target parameters & Parametrisation of the primordial matter power spectrum\vspace{3pt}\\
  $\boldsymbol{\uppsi} \in \mathbb{R}^T$ & Nuisance parameters & \makecell{Random numbers involved in the initial phase realisation,\\ instrumental noise, etc.}\vspace{3pt}\\
  $\boldsymbol{\textbf{d}} \in \mathbb{R}^D$ & Raw data & Galaxy number counts in a three-dimensional map of the survey volume\vspace{3pt}\\
  $\boldsymbol{\Phi} \in \mathbb{R}^P$ & \makecell{Summary statistics\\ of the data (observed or simulated)} & \makecell{Summaries of the galaxy number count field,\\ such as its estimated power spectrum}\vspace{3pt}\\
  $\boldsymbol{\Phi}_\mathrm{O} \in \mathbb{R}^P$ & \makecell{Summary statistics\\ of the observations} & Summaries of the observed galaxy number count field\vspace{3pt}\\
  $\boldsymbol{\Phi}_{\boldsymbol{\uptheta}} \in \mathbb{R}^P$ & \makecell{Summary statistics\\ of simulated data} & \makecell{Summaries of a galaxy number count field,\\ simulated with primordial matter power spectrum given by $\boldsymbol{\uptheta}$}\vspace{3pt}\\
  $\textbf{s} \in \mathbb{R}^P$ & Virtual signal & \makecell{True summaries of the galaxy number count field,\\ if they were not degraded by nuisances}\\
  \hline
  \end{tabular}
\caption{The statistical variables appearing in section \ref{ssec:Design of an effective likelihood for black-box models} and their interpretation in the context of galaxy survey data analysis.\label{tab:vartable}} 
\end{table*}

Table \ref{tab:vartable} provides an overview of the different variables appearing in this section and their interpretation in the context of galaxy survey data analysis.

\subsubsection{The data model}
\label{sssec:The data model}

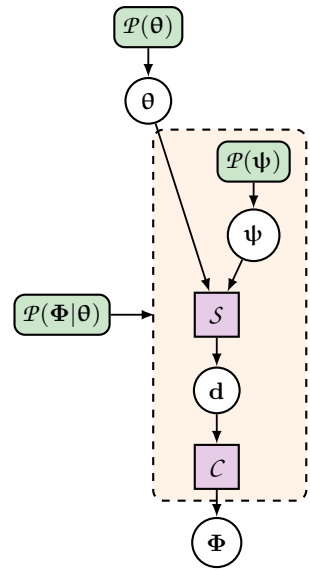
\begin{figure}
\begin{center}
\begin{tikzpicture}
	\pgfdeclarelayer{background}
	\pgfdeclarelayer{foreground}
	\pgfsetlayers{background,main,foreground}

	\tikzstyle{probability}=[draw, thick, text centered, rounded corners, minimum height=1em, minimum width=1em, fill=green!20]
	\tikzstyle{deterministic}=[draw, thick, text centered, minimum height=1.8em, minimum width=1.8em, fill=violet!20]
	\tikzstyle{variabl}=[draw, thick, text centered, circle, minimum height=1em, minimum width=1em, fill=white]
	\tikzstyle{plate} = [draw, thick, rectangle, rounded corners, dashed, fill=orange!10]

	\def\blockdist{0.7}
	\def\modeldist{2.0}

    \node (thetaproba) [probability]
    {$\p(\boldsymbol{\uptheta})$};
    \path (thetaproba.south)+(0,-\blockdist) node (theta) [variabl]
    {$\boldsymbol{\uptheta}$};
    \path [plate] (theta.south)+(0.2em,-0.2em) rectangle (3*\blockdist,-9*\blockdist)
    {};
    \path (theta.south)+(2*\blockdist,-0.7*\blockdist) node (psiproba) [probability]
    {$\p(\boldsymbol{\uppsi})$};
    \path (psiproba.south)+(0,-\blockdist) node (psi) [variabl]
    {$\boldsymbol{\uppsi}$};
    \path (psi.south)+(-0.7*\blockdist,-\blockdist) node (S) [deterministic]
    {$\mathpzc{S}$};
    \path (S.south)+(0,-\blockdist) node (d) [variabl]
    {$\textbf{d}$};
    \path (d.south)+(0,-\blockdist) node (C) [deterministic]
    {$\mathpzc{C}$};
    \path (C.south)+(0,-\blockdist) node (Phi) [variabl]
    {$\boldsymbol{\Phi}$};
    \path (S.west)+(-2.5*\blockdist,0) node (phiproba) [probability]
    {$\p(\boldsymbol{\Phi}|\boldsymbol{\uptheta})$};
    \path (phiproba.east)+(\blockdist,0) node (phiprobaB)
    {};
    
	\path [draw, line width=0.7pt, arrows={-latex}] (thetaproba) -- (theta);
	\path [draw, line width=0.7pt, arrows={-latex}] (psiproba) -- (psi);
	\path [draw, line width=0.7pt, arrows={-latex}] (theta) -- (S);
	\path [draw, line width=0.7pt, arrows={-latex}] (psi) -- (S);
	\path [draw, line width=0.7pt, arrows={-latex}] (S) -- (d);
	\path [draw, line width=0.7pt, arrows={-latex}] (d) -- (C);
	\path [draw, line width=0.7pt, arrows={-latex}] (C) -- (Phi);
	\path [draw, line width=0.7pt, arrows={-latex}] (phiproba.east) -- (phiprobaB);

\end{tikzpicture}
\end{center}
\caption{Hierarchical representation of the black-box model used in this work. The rounded green boxes represent probability distributions and the purple square represent deterministic functions. The variables are $\boldsymbol{\uptheta}$ (the target parameters), $\boldsymbol{\uppsi}$ (the nuisance parameters), $\textbf{d}$ (the full data), $\boldsymbol{\Phi}$ (the summary statistics). The orange dashed rectangle represents the data-generating process, it gives the true (unknown) likelihood $L(\boldsymbol{\uptheta})$ when $\boldsymbol{\Phi}= \boldsymbol{\Phi}_\mathrm{O}$ (the summary statistics of the observations).\label{fig:blackbox_exact}}
\end{figure}

We assume given a black-box model that provides realistic predictions for artificial observations when provided with all necessary input parameters. These consist of the target vector $\boldsymbol{\uptheta} \in \mathbb{R}^S$ parametrising the primordial power spectrum, and of nuisance parameters $\boldsymbol{\uppsi} \in \mathbb{R}^T$, independent of $\boldsymbol{\uptheta}$. Nuisance parameters account for the entire stochasticity of the data model, such as initial phases, noise realisations, sample variance, etc. In our case, nuisance parameters will be all the random numbers generated by the galaxy survey simulator for initial conditions and instrumental noise; there are typically $\mathcal{O}(10^7)$ of those. Once realisations of $\boldsymbol{\uptheta}$ and $\boldsymbol{\uppsi}$ are specified, the output of the simulation $\textbf{d} \in \mathbb{R}^D$ is a deterministic numerical function $\mathpzc{S}$, i.e. $\p(\textbf{d}|\boldsymbol{\uptheta}, \boldsymbol{\uppsi}) = \updelta_\mathrm{D}(\textbf{d} - \mathpzc{S}(\boldsymbol{\uptheta}, \boldsymbol{\uppsi}))$, where the symbol $\p$ denotes a probability distribution function (pdf) and $\updelta_\mathrm{D}$ a Dirac delta distribution. We refer to such realisations as mock observations. As usual in ABC approaches, the (often high-dimensional) raw prediction $\textbf{d}$ can be compressed to a set of summary statistics $\boldsymbol{\Phi} \in \mathbb{R}^P$. We assume that this compression is a deterministic function $\mathpzc{C}$ of $\textbf{d}$, i.e. $\p(\boldsymbol{\Phi} |\textbf{d}) = \updelta_\mathrm{D}(\boldsymbol{\Phi}-\mathpzc{C}(\textbf{d}))$. It can be included in the model, so that the black-box is $\mathpzc{B} \equiv \mathpzc{C} \circ \mathpzc{S}$ and
\begin{equation}
\p(\boldsymbol{\Phi}|\boldsymbol{\uptheta}, \boldsymbol{\uppsi}) = \updelta_\mathrm{D}(\boldsymbol{\Phi} - \mathpzc{B}(\boldsymbol{\uptheta}, \boldsymbol{\uppsi})).
\label{eq:black_box_proba}
\end{equation}
A graphical representation of the Bayesian hierarchical data model is presented in figure \ref{fig:blackbox_exact}. 

\subsubsection{The exact Bayesian problem}
\label{sssec:The exact Bayesian problem}

Denoting by $\boldsymbol{\Phi}_\mathrm{O}$ the summary statistics of the observations, the inference problem considered is
\begin{equation}
\p(\boldsymbol{\uptheta}|\boldsymbol{\Phi})_{|\boldsymbol{\Phi}=\boldsymbol{\Phi}_\mathrm{O}} = L(\boldsymbol{\uptheta}) \frac{\p(\boldsymbol{\uptheta})}{Z_{\boldsymbol{\Phi}}},
\end{equation}
where the likelihood is
\begin{equation}
L(\boldsymbol{\uptheta}) \equiv \p(\boldsymbol{\Phi}|\boldsymbol{\uptheta})_{|\boldsymbol{\Phi}=\boldsymbol{\Phi}_\mathrm{O}}
\end{equation}
and the normalisation constant is $Z_{\boldsymbol{\Phi}} \equiv \p(\boldsymbol{\Phi})_{|\boldsymbol{\Phi}=\boldsymbol{\Phi}_\mathrm{O}}$. By marginalising over $\boldsymbol{\uppsi}$ and using equation \eqref{eq:black_box_proba}, we have
\begin{eqnarray}
L(\boldsymbol{\uptheta}) & = & \int \p(\boldsymbol{\Phi}|\boldsymbol{\uptheta}, \boldsymbol{\uppsi})_{|\boldsymbol{\Phi}=\boldsymbol{\Phi}_\mathrm{O}} \p(\boldsymbol{\uppsi}) \, \mathrm{d}\boldsymbol{\uppsi} \nonumber\\
& = & \int \updelta_\mathrm{D}(\boldsymbol{\Phi}_\mathrm{O} - \mathpzc{B}(\boldsymbol{\uptheta}, \boldsymbol{\uppsi})) \p(\boldsymbol{\uppsi}) \, \mathrm{d}\boldsymbol{\uppsi}.
\label{eq:true_likelihood}
\end{eqnarray}
From equation \eqref{eq:true_likelihood}, it is clear that the likelihood involves an intractable integral, the computation of which would require exactly hitting the observed summaries $\boldsymbol{\Phi}_\mathrm{O}$ with the black-box. We are therefore not able to explicitly formulate the true likelihood distribution for the considered problem.

\subsubsection{The effective likelihood}
\label{sssec:The effective likelihood}

\begin{figure}
\begin{center}
\begin{tikzpicture}
	\pgfdeclarelayer{background}
	\pgfdeclarelayer{foreground}
	\pgfsetlayers{background,main,foreground}

	\tikzstyle{probability}=[draw, thick, text centered, rounded corners, minimum height=1em, minimum width=1em, fill=green!20]
	\tikzstyle{deterministic}=[draw, thick, text centered, minimum height=1.8em, minimum width=1.8em, fill=violet!20]
	\tikzstyle{variabl}=[draw, thick, text centered, circle, minimum height=1em, minimum width=1em, fill=white]
	\tikzstyle{plate} = [draw, thick, rectangle, rounded corners, dashed, fill=orange!10]

	\def\blockdist{0.7}
	\def\modeldist{2.0}

    \node (thetaproba) [probability]
    {$\p(\boldsymbol{\uptheta})$};
    \path (thetaproba.south)+(0,-\blockdist) node (theta) [variabl]
    {$\boldsymbol{\uptheta}$};
    \path [plate] (theta.south)+(-1.5*\blockdist,-0.6em) rectangle (1.5*\blockdist,-6.2*\blockdist)
    {};
    \path (theta.south)+(0,-\blockdist) node (sproba) [probability]
    {$\p(\textbf{s}|\boldsymbol{\uptheta})$};
    \path (sproba.south)+(0,-\blockdist) node (s) [variabl]
    {$\textbf{s}$};
    \path (s.south)+(0,-\blockdist) node (phiprobaA) [probability]
    {$\p(\boldsymbol{\Phi}|\textbf{s})$};
    \path (phiprobaA.south)+(0,-\blockdist) node (phi) [variabl]
    {$\boldsymbol{\Phi}$};
    \path (s.west)+(-2.7*\blockdist,0) node (phiprobaB) [probability]
    {$\p(\boldsymbol{\Phi}|\boldsymbol{\uptheta})$};
    \path (phiprobaB.east)+(0.9*\blockdist,0) node (phiprobaBB)
    {};
    
	\path [draw, line width=0.7pt, arrows={-latex}] (thetaproba) -- (theta);
	\path [draw, line width=0.7pt, arrows={-latex}] (theta) -- (sproba);
	\path [draw, line width=0.7pt, arrows={-latex}] (sproba) -- (s);
	\path [draw, line width=0.7pt, arrows={-latex}] (s) -- (phiprobaA);
	\path [draw, line width=0.7pt, arrows={-latex}] (phiprobaA) -- (phi);
	\path [draw, line width=0.7pt, arrows={-latex}] (phiprobaB.east) -- (phiprobaBB);

\end{tikzpicture}
\end{center}
\caption{Hierarchical representation of the approximate model used for inference of the black-box model of figure \ref{fig:blackbox_exact}: a virtual signal $\textbf{s}$ has been introduced as a latent variable. The orange dashed rectangle represents the assumption made about the data-generating process, it gives the effective likelihood $\widehat{L}^N(\boldsymbol{\uptheta})$ when conditioning on a set of simulations $\lbrace \boldsymbol{\Phi}_{\boldsymbol{\uptheta}}^{(i)} \rbrace$ and using $\boldsymbol{\Phi}= \boldsymbol{\Phi}_\mathrm{O}$.\label{fig:blackbox_effective}}
\end{figure}
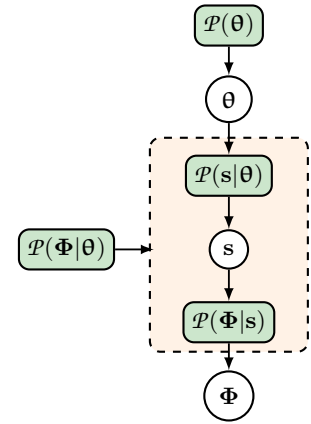

To overcome this difficulty, in this section we derive an effective likelihood that allows us to perform inference by requiring black-box model evaluations only. For every $\boldsymbol{\uptheta}$, we can generate an ensemble of $N$ mock data realisations $\boldsymbol{\Phi}_{\boldsymbol{\uptheta}}^{(i)} = \mathpzc{B}(\boldsymbol{\uptheta}, \boldsymbol{\uppsi}^{(i)})$ with $i \in  \{1, ... ,N\}$, by drawing independent and identically-distributed realisations of nuisance parameters $\boldsymbol{\uppsi}$ from the probability distribution $\p(\boldsymbol{\uppsi})$ and evaluating the black-box. The approach then consists in explicitly conditioning all probabilities on $\lbrace \boldsymbol{\Phi}_{\boldsymbol{\uptheta}}^{(i)} \rbrace$. Using Bayes' theorem, we have:
\begin{eqnarray}
\p(\boldsymbol{\uptheta} | \boldsymbol{\Phi} , \lbrace \boldsymbol{\Phi}_{\boldsymbol{\uptheta}}^{(i)} \rbrace) & = & \frac{ \p(\boldsymbol{\Phi}, \lbrace \boldsymbol{\Phi}_{\boldsymbol{\uptheta}}^{(i)} \rbrace | \boldsymbol{\uptheta}) \, \p(\boldsymbol{\uptheta})}{\p(\boldsymbol{\Phi}, \lbrace \boldsymbol{\Phi}_{\boldsymbol{\uptheta}}^{(i)} \rbrace)} \nonumber\\
& = & \frac{ \p(\boldsymbol{\Phi}, \lbrace \boldsymbol{\Phi}_{\boldsymbol{\uptheta}}^{(i)} \rbrace | \boldsymbol{\uptheta} )}{\p( \lbrace \boldsymbol{\Phi}_{\boldsymbol{\uptheta}}^{(i)} \rbrace)} \frac{\p(\boldsymbol{\uptheta})}{\p(\boldsymbol{\Phi}) }
\label{eq:bayes_approx_posterior}.
\end{eqnarray}
Using for $\boldsymbol{\Phi}$ the observed data $\boldsymbol{\Phi}_\mathrm{O}$, we thus have the new inference problem
\begin{eqnarray}
\p(\boldsymbol{\uptheta}|\boldsymbol{\Phi})_{|\boldsymbol{\Phi}=\boldsymbol{\Phi}_\mathrm{O}} & \approx & \p(\boldsymbol{\uptheta} | \boldsymbol{\Phi} , \lbrace \boldsymbol{\Phi}_{\boldsymbol{\uptheta}}^{(i)} \rbrace)_{|\boldsymbol{\Phi}=\boldsymbol{\Phi}_\mathrm{O}} \nonumber\\
& = & \widehat{L}^N(\boldsymbol{\uptheta}) \frac{\p(\boldsymbol{\uptheta})}{Z_{\boldsymbol{\Phi}}}
\label{eq:approx_inference_problem}
\end{eqnarray}
where we have defined the first factor on the right-hand side of equation \eqref{eq:bayes_approx_posterior} evaluated at $\boldsymbol{\Phi}_\mathrm{O}$ to be the effective likelihood (a computable approximation of the true likelihood):
\begin{equation}
\widehat{L}^N(\boldsymbol{\uptheta}) \equiv \frac{ \p(\boldsymbol{\Phi}, \lbrace \boldsymbol{\Phi}_{\boldsymbol{\uptheta}}^{(i)} \rbrace | \boldsymbol{\uptheta})_{|\boldsymbol{\Phi}=\boldsymbol{\Phi}_\mathrm{O}}}{\p(\lbrace \boldsymbol{\Phi}_{\boldsymbol{\uptheta}}^{(i)} \rbrace)}.
\label{eq:effective_likelihood_def}
\end{equation}

To arrive at a more explicit expression for the effective likelihood, we assume that observed data $\boldsymbol{\Phi}_\mathrm{O}$ and mock realisations $\boldsymbol{\Phi}_{\boldsymbol{\uptheta}}^{(i)}$ are drawn from a common  (but unknown)  virtual signal $\textbf{s} \in \mathbb{R}^P$. The hierarchical representation of this effective data model is presented in figure \ref{fig:blackbox_effective}. We assume that $\textbf{s}$ carries the deterministic information on the target parameters $\boldsymbol{\uptheta}$, which is common to the data and to the mock observations; therefore, $\boldsymbol{\Phi}_\mathrm{O}$ and $\boldsymbol{\Phi}_{\boldsymbol{\uptheta}}^{(i)}$ only differ by stochastic uncertainties described by the nuisance parameters, which carry no information of interest. Intuitively, $\textbf{s}$ represents the ``true'' version of the summaries $\boldsymbol{\Phi}$, which is degraded by nuisances $\boldsymbol{\uppsi}$. Under this assumption, we introduce the probability distribution $\p(\boldsymbol{\Phi}|\textbf{s})$, from which data realisations are drawn independently once the virtual signal $\textbf{s}$ is given. Since we do not want to infer $\textbf{s}$ explicitly, marginalisation yields:
\begin{eqnarray}
\p(\boldsymbol{\Phi}, \lbrace \boldsymbol{\Phi}_{\boldsymbol{\uptheta}}^{(i)} \rbrace | \boldsymbol{\uptheta}) & = & \int \p(\boldsymbol{\Phi}, \lbrace \boldsymbol{\Phi}_{\boldsymbol{\uptheta}}^{(i)} \rbrace, \textbf{s} | \boldsymbol{\uptheta} ) \, \mathrm{d} \textbf{s} \\
& = & \int \p(\boldsymbol{\Phi}|\textbf{s}) \p(\lbrace \boldsymbol{\Phi}_{\boldsymbol{\uptheta}}^{(i)} \rbrace | \textbf{s}) \p(\textbf{s}|\boldsymbol{\uptheta}) \, \mathrm{d} \textbf{s}. \nonumber
\end{eqnarray}
In some models, it may be possible to derive $\textbf{s}$ from the model parameters $\boldsymbol{\uptheta}$. However, this is generally not true for numerical simulators, where the expected summaries have to be estimated through averaging mock realisations. In this work, in absence of prior information on $\textbf{s}$, we set $\p(\textbf{s}|\boldsymbol{\uptheta})$ to a constant. The joint distribution of mock observations for a given virtual signal factorises, so that one can write
\begin{equation}
\p(\boldsymbol{\Phi}, \lbrace \boldsymbol{\Phi}_{\boldsymbol{\uptheta}}^{(i)} \rbrace | \boldsymbol{\uptheta}) \propto \int \p(\boldsymbol{\Phi}|\textbf{s}) \left[\prod_{n=1}^N \p(\boldsymbol{\Phi}_{\boldsymbol{\uptheta}}^{(i)}| \textbf{s})\right] \mathrm{d} \textbf{s}.
\label{eq:marginalisation}
\end{equation}
In order to marginalise over $\textbf{s}$, we need to postulate a parametric form for the pdf $\p(\boldsymbol{\Phi}|\textbf{s})$. In this work, we assume a Gaussian distribution centred on the virtual signal $\textbf{s}$ with a covariance matrix $\boldsymbol{\Sigma}_{\boldsymbol{\uptheta}}$ quantifying the stochastic uncertainties inherent to the observations,\footnote{The investigation of different choices for $\p(\boldsymbol{\Phi}|\textbf{s}) $ is left to future investigations.}
\begin{equation}
-2 \log \p(\boldsymbol{\Phi}|\textbf{s}) = \log\left| 2\pi\boldsymbol{\Sigma}_{\boldsymbol{\uptheta}} \right| + (\boldsymbol{\Phi}-\textbf{s})^\intercal \boldsymbol{\Sigma}_{\boldsymbol{\uptheta}}^{-1} (\boldsymbol{\Phi}-\textbf{s}).
\label{eq:virtual_signal_pdf}
\end{equation}
The virtual signal can therefore be interpreted as the expectation $\mathrm{E}$ of $\boldsymbol{\Phi}_{\boldsymbol{\uptheta}}$ once $\boldsymbol{\uptheta}$ is specified: $\textbf{s} = \mathrm{E}\left[ \boldsymbol{\Phi}_{\boldsymbol{\uptheta}} \right]$. We also have $\boldsymbol{\Sigma}_{\boldsymbol{\uptheta}} = \mathrm{E}\left[ (\boldsymbol{\Phi}_{\boldsymbol{\uptheta}}-\textbf{s})(\boldsymbol{\Phi}_{\boldsymbol{\uptheta}}-\textbf{s})^\intercal \right]$.

Under these assumptions, marginalisation over $\textbf{s}$ gives the effective likelihood as $\widehat{L}^N(\boldsymbol{\uptheta}) = \exp\left[ \hat{\ell}^N(\boldsymbol{\uptheta}) \right]$ (the details of the computation are given in appendix \ref{apx:Derivation of the effective likelihood}), with:
\begin{equation}
-2 \hat{\ell}^N(\boldsymbol{\uptheta}) = \log\left| 2\pi\boldsymbol{\hat{\Sigma}}_{\boldsymbol{\uptheta}}' \right| + (\boldsymbol{\Phi}_\mathrm{O} - \boldsymbol{\hat{\Phi}}_{\boldsymbol{\uptheta}})^\intercal \boldsymbol{\hat{\Sigma}}_{\boldsymbol{\uptheta}}'^{-1} (\boldsymbol{\Phi}_\mathrm{O} - \boldsymbol{\hat{\Phi}}_{\boldsymbol{\uptheta}}),
\label{eq:effective_likelihood}
\end{equation}
where
\begin{equation}
\boldsymbol{\hat{\Phi}}_{\boldsymbol{\uptheta}} \equiv \mathrm{E}^N \left[ \boldsymbol{\Phi}_{\boldsymbol{\uptheta}} \right] = \frac{1}{N}\sum_{i=1}^{N} \boldsymbol{\Phi}_{\boldsymbol{\uptheta}}^{(i)}
\label{eq:estimated_mean}
\end{equation}
is the ensemble mean of mock observations and $\mathrm{E}^N$ stands for the empirical average over the set. The covariance matrix of $\widehat{L}^N(\boldsymbol{\uptheta})$ and its inverse are defined by 
\begin{equation}
\boldsymbol{\hat{\Sigma}}_{\boldsymbol{\uptheta}}' \equiv \frac{N+1}{N}\boldsymbol{\hat{\Sigma}}_{\boldsymbol{\uptheta}}, \quad \boldsymbol{\hat{\Sigma}}_{\boldsymbol{\uptheta}}'^{-1} \equiv \left(\frac{N+1}{N}\right)^{-1} \boldsymbol{\hat{\Sigma}}_{\boldsymbol{\uptheta}}^{-1},
\label{eq:covariance_effective_likelihood}
\end{equation}
where given a sufficiently large number of mock observations, a computable approximation of $\boldsymbol{\Sigma}_{\boldsymbol{\uptheta}}$ is estimated in the following way:
\begin{eqnarray}
\boldsymbol{\Sigma}_{\boldsymbol{\uptheta}} \approx \boldsymbol{\hat{\Sigma}}_{\boldsymbol{\uptheta}} & \equiv & \mathrm{E}^N\left[ (\boldsymbol{\Phi}_{\boldsymbol{\uptheta}} - \boldsymbol{\hat{\Phi}}_{\boldsymbol{\uptheta}})(\boldsymbol{\Phi}_{\boldsymbol{\uptheta}} - \boldsymbol{\hat{\Phi}}_{\boldsymbol{\uptheta}})^\intercal \right] \nonumber\\
& = & \frac{1}{N-1} \sum_{i=1}^N (\boldsymbol{\Phi}_{\boldsymbol{\uptheta}}^{(i)} - \boldsymbol{\hat{\Phi}}_{\boldsymbol{\uptheta}}) (\boldsymbol{\Phi}_{\boldsymbol{\uptheta}}^{(i)} - \boldsymbol{\hat{\Phi}}_{\boldsymbol{\uptheta}})^\intercal .
\label{eq:estimated_covariance}
\end{eqnarray}
The effective likelihood also requires an estimator $\boldsymbol{\hat{\Sigma}}_{\boldsymbol{\uptheta}}^{-1}$ of the inverse covariance matrix $\boldsymbol{\Sigma}_{\boldsymbol{\uptheta}}^{-1}$ (see equation \eqref{eq:covariance_effective_likelihood}). As argued by \citet{SellentinHeavens2016} and \citet{JeffreyAbdalla2018}, the proper Bayesian treatment would in fact consist in replacing the Gaussian likelihood by an alternative, corrected distribution. Nevertheless, keeping a Gaussian effective likelihood is an essential requirement of the present ABC technique; we will therefore be content with the \citet{Hartlap2007} correction, which consists of replacing the true inverse covariance matrix by a scaled inverse sample covariance matrix:
\begin{equation}
\boldsymbol{\Sigma}_{\boldsymbol{\uptheta}}^{-1} \approx \boldsymbol{\hat{\Sigma}}_{\boldsymbol{\uptheta}}^{-1} \equiv \alpha \left( \boldsymbol{\hat{\Sigma}}_{\boldsymbol{\uptheta}} \right)^{-1}, \quad \alpha \equiv \frac{N-P-2}{N-1},
\label{eq:estimated_inverse_covariance}
\end{equation}
where $P$ is the number of summary statistics. This correction debiases the expectation value $\left\langle \left( \boldsymbol{\hat{\Sigma}}_{\boldsymbol{\uptheta}} \right)^{-1} \right\rangle = \alpha^{-1} \boldsymbol{\Sigma}_{\boldsymbol{\uptheta}}^{-1}$ of the estimator, under the assumption that $\boldsymbol{\Sigma}_{\boldsymbol{\uptheta}}^{-1}$ is inverse-Wishart distributed.

The limiting approximation of $\widehat{L}^N(\boldsymbol{\uptheta})$ when $N \rightarrow \infty$ is $\widetilde{L}(\boldsymbol{\uptheta}) = \p(\boldsymbol{\Phi}|\textbf{s})_{|\boldsymbol{\Phi} = \boldsymbol{\Phi}_{\mathrm{O}}} = \exp \left[ \tilde{\ell}(\boldsymbol{\uptheta}) \right]$, with (as intended)
\begin{equation}
-2 \tilde{\ell}(\boldsymbol{\uptheta}) \equiv \log\left| 2\pi \boldsymbol{\Sigma}_{\boldsymbol{\uptheta}} \right| + (\boldsymbol{\Phi}_\mathrm{O}-\textbf{s})^\intercal \boldsymbol{\Sigma}_{\boldsymbol{\uptheta}}^{-1} (\boldsymbol{\Phi}_\mathrm{O}-\textbf{s}).
\label{eq:effective_likelihood_limiting}
\end{equation}

There are a number of interesting similarities and differences between the effective likelihood (equation \eqref{eq:effective_likelihood}) and the virtual signal pdf (equation \eqref{eq:effective_likelihood_limiting}). First, both are Gaussian distributions with respect to the observed data $\boldsymbol{\Phi}_\mathrm{O}$. Second, the mean of the effective likelihood is given by the empirical average of simulated summaries, $\boldsymbol{\hat{\Phi}}_{\boldsymbol{\uptheta}} = \mathrm{E}^N \left[ \boldsymbol{\Phi}_{\boldsymbol{\uptheta}} \right]$. This is an unbiased estimator of the virtual signal $\textbf{s} = \mathrm{E}\left[ \boldsymbol{\Phi}_{\boldsymbol{\uptheta}} \right]$ for a sufficiently large number of simulations. Importantly, the replacement of the expectation $\mathrm{E}$ by an empirical average $\mathrm{E}^N$ was not an assumption (contrary to the synthetic likelihood, \citealp{Wood2010,Price2018}), but naturally appeared in the derivation (see appendix \ref{apx:Derivation of the effective likelihood}). Finally, the covariance of the effective likelihood is $\frac{N+1}{N} \boldsymbol{\hat{\Sigma}}_{\boldsymbol{\uptheta}}$, a multiple of the estimated covariance $\boldsymbol{\hat{\Sigma}}_{\boldsymbol{\uptheta}}$. The numerical prefactor, similar to Bessel's correction for the empirical sample variance, can be understood as follows. For a small number of simulations, the latent space associated to the virtual signal increases the observed scatter. For example, for $N=1$, the covariance to be used in equation \eqref{eq:effective_likelihood} is $2\boldsymbol{\hat{\Sigma}}_{\boldsymbol{\uptheta}}$, which reflects the fact that observed and simulated data $\boldsymbol{\Phi}_\mathrm{O}$ and $\boldsymbol{\Phi}_{\boldsymbol{\uptheta}}$ can be drawn from opposite ends of the scatter around $\textbf{s}$ of $\p(\boldsymbol{\Phi}|\textbf{s})$. However, when $N \rightarrow \infty$, $\frac{N+1}{N} \boldsymbol{\hat{\Sigma}}_{\boldsymbol{\uptheta}} \longrightarrow \boldsymbol{\Sigma}_{\boldsymbol{\uptheta}}$, the intrinsic covariance of $\p(\boldsymbol{\Phi}|\textbf{s})$. This result is reasonable, since for large sample sizes the ensemble mean $\boldsymbol{\hat{\Phi}}_{\boldsymbol{\uptheta}}$ converges to $\textbf{s}$, the mean of $\p(\boldsymbol{\Phi}|\textbf{s})$; in the same limit, we expect the covariance of the effective likelihood to approach $\boldsymbol{\Sigma}_{\boldsymbol{\uptheta}}$, the covariance of $\p(\boldsymbol{\Phi}|\textbf{s})$.

Note that the scaling factor $\frac{N+1}{N}$ directly arises from the presence of the virtual signal, as shown in appendix \ref{apx:Derivation of the effective likelihood}. Alternatively, we could have directly assumed a parametric form for the pdf $\p(\boldsymbol{\Phi}|\boldsymbol{\uptheta})$ without introducing the latent variable $\textbf{s}$. For instance, the synthetic likelihood directly amounts to postulating equation \eqref{eq:effective_likelihood}, but without the scaling factor $\frac{N+1}{N}$ for the estimated covariance matrix. This means that the end result for $\hat{\ell}^N(\boldsymbol{\uptheta})$ has little sensitivity to our treatment and assumption for $\p(\textbf{s}|\boldsymbol{\uptheta})$. Furthermore, since $\frac{N+1}{N} \longrightarrow 1$ when $N \rightarrow \infty$, our result with the virtual signal and the synthetic likelihood are equivalent, provided that the number of simulations is large enough.

Determining $\boldsymbol{\hat{\Phi}}_{\boldsymbol{\uptheta}}$ and $\boldsymbol{\hat{\Sigma}}_{\boldsymbol{\uptheta}}$ requires $N$ model evaluations per target parameters $\boldsymbol{\uptheta}$. In the next section, we show how the evaluation of this effective likelihood can be made more efficient when exploiting prior information on $\boldsymbol{\uptheta}$.

\subsection{Linearisation of black-box models}
\label{ssec:Linearisation of black-box models}

The numerical cost of the computation of the effective likelihood described in section \ref{ssec:Design of an effective likelihood for black-box models} may be prohibitively large when the full parameter space has to be explored. However, such an extensive exploration is not always required: often, one has sufficient information to be only interested in a small region of parameter space around a specific prediction. This requirement is fulfilled for inferences of the primordial power spectrum from galaxy surveys. The target parameters $\boldsymbol{\uptheta}$ typically consist of power spectrum amplitudes in about a hundred different bands of wavevectors; thus the corresponding parameter space is very large. However, the CMB already provides exquisite measurements of the primordial cosmological power spectrum \citep[e.g.][]{Planck2018I}. Any large deviations from these previous measurements would most likely be rejected on methodological grounds (sample size, uncontrolled systematics, etc.) and not be attributed to new physics. It therefore seems reasonable to focus the inference of the primordial power spectrum from galaxy surveys within a narrow region in parameter space around a previous estimate obtained from CMB results.

Following this reasoning, we focus on searching for solutions corresponding to small deviations $\Delta \boldsymbol{\uptheta}$ around an expansion point $\boldsymbol{\uptheta}_0$. Consequently, the target parameters are given by $\boldsymbol{\uptheta} = \boldsymbol{\uptheta}_0 + \Delta \boldsymbol{\uptheta}$. Assuming that $\boldsymbol{\hat{\Phi}}_{\boldsymbol{\uptheta}}$ is differentiable with respect to $\boldsymbol{\uptheta}$, we perform a first-order Taylor expansion in $\Delta \boldsymbol{\uptheta}$ around $\boldsymbol{\uptheta}_0$:
\begin{equation}
\boldsymbol{\hat{\Phi}}_{\boldsymbol{\uptheta}} \approx \textbf{f}_0 + \nabla \textbf{f}_0 \cdot (\boldsymbol{\uptheta}-\boldsymbol{\uptheta}_0) \equiv \textbf{f}(\boldsymbol{\uptheta}) ,
\label{eq:linearised_black_box}
\end{equation}
where the defined function $\textbf{f}$ is a linearised version of the averaged black-box. The first term corresponds to the mean mock observations at the expansion point $\boldsymbol{\uptheta}_0$, i.e. $\textbf{f}_0 \equiv \boldsymbol{\hat{\Phi}}_{\boldsymbol{\uptheta}_0}$, while the second term involves $\nabla \textbf{f}_0$, a $P \times  S$ matrix corresponding to the gradient of mean mock observations at $\boldsymbol{\uptheta}_0$, whose components are $\left(\nabla \textbf{f}_0\right)_{ps} \equiv \frac{\partial \boldsymbol{\hat{\Phi}}_{\boldsymbol{\uptheta}_0 p}}{\partial \boldsymbol{\uptheta}_s}$. In this work, we estimate the gradient $\nabla \textbf{f}_0$ via finite differencing: given a small step size $h$, each column is approximated by
\begin{equation}
\left(\nabla \textbf{f}_0\right)^\intercal_s \approx \frac{\textbf{f}(\boldsymbol{\uptheta}_s)-\textbf{f}_0}{h}, \quad \mathrm{with} \quad \boldsymbol{\uptheta}_s \equiv \boldsymbol{\uptheta}_0 + h \, (\updelta_\mathrm{K}^{ss'})_{s' \in \llbracket 1, S\rrbracket},
\label{eq:gradient_finite_differencing}
\end{equation}
where $\updelta_\mathrm{K}$ is the Kronecker delta. To avoid obtaining a noisy gradient, nuisance parameters $\boldsymbol{\uppsi}$ are kept at the same values in the simulations used to compute the $\textbf{f}(\boldsymbol{\uptheta}_s)$ and $\textbf{f}_0$. Further, we neglect the dependence of the covariance matrix on $\boldsymbol{\uptheta}$ and we estimate it at the expansion point according to equation \eqref{eq:estimated_covariance}, i.e. $\boldsymbol{\hat{\Sigma}}_{\boldsymbol{\uptheta}}' \approx \boldsymbol{\hat{\Sigma}}_{\boldsymbol{\uptheta}_0}' \equiv \textbf{C}_0$. Importantly, fully characterising $\textbf{f}$ under these assumptions only requires evaluations of the simulator, thus ensuring that the data model remains a black-box.

Using the linearised data model $\textbf{f}$ as a proxy for the data model described in section \ref{sssec:The effective likelihood} simplifies the expression of the effective likelihood (equation \eqref{eq:effective_likelihood}) to
\begin{equation}
-2 \hat{\ell}^N(\boldsymbol{\uptheta}) \approx \log\left| 2\pi\textbf{C}_0 \right| + \left[\boldsymbol{\Phi}_\mathrm{O} - \textbf{f}(\boldsymbol{\uptheta})\right]^\intercal \textbf{C}_0^{-1} \left[\boldsymbol{\Phi}_\mathrm{O} - \textbf{f}(\boldsymbol{\uptheta})\right].
\label{eq:linearised_effective_likelihood}
\end{equation}

Evaluating $\textbf{f}_0$ and $\textbf{C}_0$ requires $N_0$ model evaluations at the expansion point $\boldsymbol{\uptheta}_0$. The computation of the gradient further requires $N_s \times S$ model evaluations. Therefore, the linearised data model is fully characterised by a fixed total of $N_0 + N_s \times S$ model evaluations. $N_0$ and $N_s$ are a user choice, but should be of the order of the dimensionality of the data space $P$. Particularly at the expansion point $\boldsymbol{\uptheta}_0$, a minimum would be $N_0 \geq P+3$ but it can be worth investing more simulations, in order for the estimated covariance matrix $\textbf{C}_0$ and its inverse to be precise. Since all nuisance parameters are kept fixed in the computation of the $\textbf{f}(\boldsymbol{\uptheta}_s)$, $N_s$ can in principle be smaller than $P$ (and as small as $1$); however $N_s \gtrsim P$ will yield a safer evaluation of $\nabla \textbf{f}_0$.

It is important to note that all required model evaluations are done once and for all. Once the linearised data model is known, it is not necessary to perform additional black-box evaluations in order to perform inference from new data. This feature makes the present approach similar to supervised machine learning algorithms, which are only trained once before being applied to multiple data sets. Furthermore, all data model evaluations can be done in parallel, or even on different machines, making the approach very suitable for grid computing.

\subsection{The power spectrum prior distribution}
\label{ssec:The power spectrum prior distribution}

In this paper, we aim at inferring the primordial matter power spectrum $P(k)$, which is a continuous function of wavenumber $k$. We parametrise it by its amplitudes at a sufficient number $S$ of support wavenumbers $k_s$. As we are particularly interested in BAOs, we fix a ``wiggle-less'' power spectrum $P_0(k)$ and we work with the ``wiggle function'' $\theta(k) \equiv P(k)/P_0(k)$ as target function. Formally, with $\textbf{P}_0 \in \mathbb{R}^S$ the vector of components $(\textbf{P}_0)_s \equiv P_0(k_s)$, the inference variable $\boldsymbol{\uptheta}$ is defined as the $S$-dimensional vector of components $P(k_s)/(\textbf{P}_0)_s$.

In order to set up the Bayesian problem, we have to formulate the prior probability distribution of $\boldsymbol{\uptheta}$. In this work, we include into our prior the following assumptions:
\begin{enumerate}
\item the power spectrum is Gaussian-distributed,
\item it is strongly constrained to live close to $P_0$,
\item it is a smooth function of wavenumber,
\item and the power spectrum $P_0$ is subject to cosmic variance.
\end{enumerate}
It follows from assumptions 1 and 2 that $\p(\boldsymbol{\uptheta})$ shall be a Gaussian distribution with mean $\boldsymbol{\uptheta}_0 \equiv \boldsymbol{1}_{\mathbb{R}^S}$, which is also to be used as the expansion point.

We now discuss assumptions 3 and 4, in order to build the covariance matrix $\textbf{S}$ of $\p(\boldsymbol{\uptheta})$. Let us define the matrix $\textbf{K}$ as a radial basis function, i.e. by its coefficients
\begin{equation}
\left(\textbf{K}\right)_{ss'} \equiv \mathrm{exp} \left[ -\frac{1}{2} \left(\frac{k_s - k_{s'}}{k_\mathrm{corr}}\right)^2 \right].
\end{equation}
The hyperparameter $k_\mathrm{corr}$ determines the length scale on which power spectrum amplitudes of different wavenumber correlate with each other. A large value of $k_\mathrm{corr}$ corresponds to a strong correlation between $\boldsymbol{\uptheta}_s$ and $\boldsymbol{\uptheta}_{s'}$, even if their corresponding wavenumbers are far from each other. On the other hand, for $k_\mathrm{corr} \ll k_s, k_{s'}$, $\textbf{K}$ becomes the identity matrix and $\boldsymbol{\uptheta}_s$ and $\boldsymbol{\uptheta}_{s'}$ do not correlate with each other at all. The previous discussion implies that the \textit{a priori} smoothness of the wiggle function can be changed by tuning $k_\mathrm{corr}$. In a realistic scenario, the prior covariance is not scale-independent as is $\textbf{K}$ ($(\textbf{K})_{ss} = 1$ for all $s$). We rather want the standard deviations $(\textbf{S})^{1/2}_{ss}$ to account for the cosmic variance affecting the power spectrum $P_0$ (and hence the mean $\boldsymbol{\uptheta}_0$). In terms of power spectrum amplitudes, cosmic variance at a scale $k$ is given by $P(k)^2/N_k$, where $N_k \propto k^3$ is the number of modes of wavenumber $k$ in the considered cosmological volume. Thus, in order to account for cosmic variance in terms of $\boldsymbol{\uptheta}$, the coefficients $\textbf{K}$ shall be multiplied by the coefficients of $\textbf{u}\textbf{u}^\intercal$, where
\begin{equation}
(\textbf{u})_s \equiv 1 + \sigma_s = 1+\frac{\alpha_\mathrm{cv}}{k_s^{3/2}}
\end{equation}
and $\alpha_\mathrm{cv}$ is a hyperparameter characterising the ``strength'' of cosmic variance given the considered volume. Finally, the amplitude of the covariance matrix $\textbf{S}$ can be captured by an overall scaling $\theta_\mathrm{norm}^2$. The final expression for the prior covariance matrix is therefore
\begin{equation}
\textbf{S} \equiv \theta_\mathrm{norm}^2 \, \textbf{u}\textbf{u}^\intercal \circ \textbf{K},
\label{eq:prior_covariance}
\end{equation}
where $\circ$ is the Hadamard product. The standard deviations on the diagonal are $(\textbf{S})^{1/2}_{ss} = \theta_\mathrm{norm} (1 + \sigma_s)$, as intended.

The resulting prior on $\boldsymbol{\uptheta}$ is characterised by a set of three hyperparameters $\{k_\mathrm{corr}, \alpha_\mathrm{cv}, \theta_\mathrm{norm}\}$, and given as
\begin{equation}
-2\log \p(\boldsymbol{\uptheta}) \equiv \log\left| 2\pi \textbf{S} \right| + (\boldsymbol{\uptheta}-\boldsymbol{\uptheta}_0)^\intercal \textbf{S}^{-1} (\boldsymbol{\uptheta} - \boldsymbol{\uptheta}_0) .
\label{eq:prior}
\end{equation}
As demonstrated in section \ref{sec:Results}, a prior of this form results in a smooth posterior mean for the primordial matter power spectrum, while incorporating reasonable uncertainties around the expansion point.

\subsection{The power spectrum effective posterior distribution}
\label{ssec:The power spectrum effective posterior distribution}

\begin{table*}
\centering
  \begin{tabular}{c c c c c c}
  \hline 
  Symbol & Meaning & Interpretation\\\hline
  $\boldsymbol{\uptheta}_0 \in \mathbb{R}^S$ & \makecell{Expansion point of the simulator\\ in parameter space and prior mean} & Fiducial primordial matter power spectrum\vspace{3pt}\\
  $\textbf{S} \in \mathbb{R}^{S \times S}$ & Prior covariance matrix & \makecell{Prior covariance matrix of the primordial matter power spectrum}\vspace{3pt}\\
  $\boldsymbol{\Phi}_\mathrm{O} \in \mathbb{R}^P$ & \makecell{Summary statistics\\ of the observations} & Summaries of the observed galaxy number count field\vspace{3pt}\\
  $\textbf{f}_0 \in \mathbb{R}^P$ & \makecell{Estimated average black-box\\ at the expansion point} & \makecell{Estimated average of summaries of simulated galaxy  fields\\ with fiducial primordial matter power spectrum}\vspace{3pt}\\
  $\textbf{C}_0 \in \mathbb{R}^{P \times P}$ & \makecell{Estimated covariance matrix of the\\ black-box at the expansion point} & \makecell{Estimated covariance of summaries of simulated galaxy fields\\ with fiducial primordial matter power spectrum}\vspace{3pt}\\
  $\nabla \textbf{f}_0 \in \mathbb{R}^{P \times S}$ & \makecell{Estimated gradient of the average\\ black-box at the expansion point} & \makecell{Estimated gradient of summaries of simulated galaxy fields\\ around the fiducial primordial matter power spectrum}\vspace{3pt}\\
  $\boldsymbol{\upgamma} \in \mathbb{R}^S$ & Posterior mean & \makecell{Reconstructed primordial matter power spectrum,\\ given the observed summaries $\boldsymbol{\Phi}_\mathrm{O}$}\vspace{3pt}\\
  $\boldsymbol{\Gamma} \in \mathbb{R}^{S \times S}$ & Posterior covariance matrix & \makecell{Uncertainties on the reconstruction of the primordial matter\\ power spectrum, given the observed summaries $\boldsymbol{\Phi}_\mathrm{O}$}\\
  \hline
  \end{tabular}
\caption{The statistical variables appearing in {\selfi} and their interpretation in the context of galaxy survey data analysis.\label{tab:varselfi}} 
\end{table*}

Using the effective likelihood with the linearised black-box data model, given in equation \eqref{eq:linearised_effective_likelihood}, and the prior given in equation \eqref{eq:prior}, we arrive at the final expression for the effective posterior distribution (see equation \eqref{eq:approx_inference_problem}). It is a Gaussian distribution,
\begin{equation}
-2\log \p(\boldsymbol{\uptheta}|\boldsymbol{\Phi})_{|\boldsymbol{\Phi}=\boldsymbol{\Phi}_\mathrm{O}} = \log \left| 2\pi\boldsymbol{\Gamma} \right| + (\boldsymbol{\uptheta}-\boldsymbol{\upgamma})^\intercal \boldsymbol{\Gamma}^{-1} (\boldsymbol{\uptheta}-\boldsymbol{\upgamma}),
\end{equation}
with mean
\begin{equation}
\boxed{\boldsymbol{\upgamma} \equiv \boldsymbol{\uptheta}_0 + \boldsymbol{\Gamma} \, (\nabla \textbf{f}_0)^\intercal \, \textbf{C}_0^{-1} (\boldsymbol{\Phi}_\mathrm{O}-\textbf{f}_0),}
\label{eq:filter_mean}
\end{equation}
and covariance matrix
\begin{equation}
\label{eq:filter_var}
\boxed{\boldsymbol{\Gamma} \equiv \left[ (\nabla \textbf{f}_0)^\intercal \, \textbf{C}_0^{-1} \nabla \textbf{f}_0 + \textbf{S}^{-1} \right]^{-1}.}
\end{equation}
The proof of this result (detailed in appendix \ref{apx:Derivation of the effective posterior}) uses the same algebra as the derivation of the Wiener filter \citep{Wiener1964}. Equations \eqref{eq:filter_mean} and \eqref{eq:filter_var} are the main result of this work. Equation \eqref{eq:filter_mean} provides a simple ``filter equation'' to infer the primordial matter power spectrum from galaxy observations via complex black-box simulations. Corresponding uncertainties are quantified by the covariance matrix given in equation \eqref{eq:filter_var}. Table \ref{tab:varselfi} summarises the variables appearing in equations \eqref{eq:filter_mean} and \eqref{eq:filter_var} and their interpretation in the context of galaxy survey data analysis.

\subsection{Optimisation of the prior hyperparameters}
\label{ssec:Optimisation of the prior hyperparameters}

As discussed in section \ref{ssec:The power spectrum prior distribution}, the chosen prior distribution involves three hyperparameters $\{k_\mathrm{corr}, \alpha_\mathrm{cv}, \theta_\mathrm{norm}\}$. By definition, $\alpha_\mathrm{cv}$ characterises the strength of cosmic variance in the considered cosmological volume, such that the number of modes at a given scale $k$ is $N_k = k^3/\alpha_\mathrm{cv}^2$. $\alpha_\mathrm{cv}$ can therefore simply be measured for a given simulator setup (i.e. box size and mesh). On the contrary, $k_\mathrm{corr}$ and $\theta_\mathrm{norm}$ are free hyperparameters and the power spectrum reconstruction ($\boldsymbol{\upgamma}$ and $\boldsymbol{\Gamma}$) generally depends on their values. In this section, we propose a procedure to find optimal values for $k_\mathrm{corr}$ and $\theta_\mathrm{norm}$.

$k_\mathrm{corr}$ indicates the \textit{a priori} smoothness of reconstructed wiggle functions $\theta(k)$ and $\theta_\mathrm{norm}$ how much they can deviate from the expansion point. Together, these two parameters characterise the functional shapes of allowed target functions (much like kernels and their hyperparameters in techniques such as Gaussian process regression, see e.g. \citealp{RasmussenWilliams2006}). In analogy with hyperparameter optimisation in machine learning, we propose to optimise $k_\mathrm{corr}$ and $\theta_\mathrm{norm}$ to reproduce the shape of a fiducial wiggle function $\boldsymbol{\uptheta}_\mathrm{fid}$, using as likelihood the effective posterior distribution derived in section \ref{ssec:The power spectrum effective posterior distribution}. More precisely, the likelihood for $k_\mathrm{corr}$ and $\theta_\mathrm{norm}$ is defined by
\begin{equation}
\begin{split}
-2\log \p(k_\mathrm{corr},\theta_\mathrm{norm}|\boldsymbol{\uptheta}_\mathrm{fid}) \equiv & \log \left| 2\pi\boldsymbol{\Gamma} \right| \\
& + (\boldsymbol{\uptheta}_\mathrm{fid}-\boldsymbol{\upgamma})^\intercal \boldsymbol{\Gamma}^{-1} (\boldsymbol{\uptheta}_\mathrm{fid}-\boldsymbol{\upgamma}),
\end{split}
\label{eq:likelihood_hyperparameters}
\end{equation}
where $\boldsymbol{\upgamma}$ and $\boldsymbol{\Gamma}$, defined by equations \eqref{eq:filter_mean} and \eqref{eq:filter_var}, are functions of $k_\mathrm{corr}$ and $\theta_\mathrm{norm}$ (through $\textbf{S}$). At this point, it is of course possible to include a hyperprior on $(k_\mathrm{corr},\theta_\mathrm{norm})$, if desired. The maximum likelihood estimator (or maximum \textit{a posteriori} estimator) then provides the optimal values of $k_\mathrm{corr}$ and $\theta_\mathrm{norm}$ to be used to infer $\boldsymbol{\uptheta}$, its functional shape being assumed to be that of $\boldsymbol{\uptheta}_\mathrm{fid}$.

Note that evaluating the likelihood given in equation \eqref{eq:likelihood_hyperparameters} is cheap once the linearised black-box $\textbf{f}$ has been computed: no additional data model evaluation is required, only low-dimensional matrix operations. This allows the optimal values of $k_\mathrm{corr}$ and $\theta_\mathrm{norm}$ to be found using standard optimisers.

\subsection{From the power spectrum to cosmological parameters}
\label{ssec:From the power spectrum to cosmological parameters}

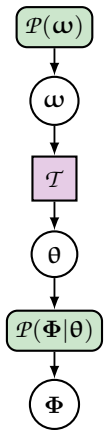
\begin{figure}
\begin{center}
\begin{tikzpicture}
	\pgfdeclarelayer{background}
	\pgfdeclarelayer{foreground}
	\pgfsetlayers{background,main,foreground}

	\tikzstyle{probability}=[draw, thick, text centered, rounded corners, minimum height=1em, minimum width=1em, fill=green!20]
	\tikzstyle{deterministic}=[draw, thick, text centered, minimum height=1.8em, minimum width=1.8em, fill=violet!20]
	\tikzstyle{variabl}=[draw, thick, text centered, circle, minimum height=1em, minimum width=1em, fill=white]

	\def\blockdist{0.7}
	\def\modeldist{2.0}

    \node (omageaproba) [probability]
    {$\p(\boldsymbol{\upomega})$};
    \path (omageaproba.south)+(0,-\blockdist) node (omega) [variabl]
    {$\boldsymbol{\upomega}$};
    \path (omega.south)+(0,-\blockdist) node (thetaproba) [deterministic]
    {$\mathpzc{T}$};
    \path (thetaproba.south)+(0,-\blockdist) node (theta) [variabl]
    {$\boldsymbol{\uptheta}$};
    \path (theta.south)+(0,-\blockdist) node (phiproba) [probability]
    {$\p(\boldsymbol{\Phi}|\boldsymbol{\uptheta})$};
    \path (phiproba.south)+(0,-\blockdist) node (phi) [variabl]
    {$\boldsymbol{\Phi}$};
    
	\path [draw, line width=0.7pt, arrows={-latex}] (omageaproba) -- (omega);
	\path [draw, line width=0.7pt, arrows={-latex}] (omega) -- (thetaproba);
	\path [draw, line width=0.7pt, arrows={-latex}] (thetaproba) -- (theta);
	\path [draw, line width=0.7pt, arrows={-latex}] (theta) -- (phiproba);
	\path [draw, line width=0.7pt, arrows={-latex}] (phiproba) -- (phi);

\end{tikzpicture}
\end{center}
\caption{Hierarchical representation of the Bayesian model for inference of cosmological parameters $\boldsymbol{\upomega}$: a layer has been added above the primordial power spectrum $\boldsymbol{\uptheta}$. The purple square represents the deterministic process generating $\boldsymbol{\uptheta}$ given $\boldsymbol{\upomega}$.\label{fig:BHM_cosmology}}
\end{figure}

As argued in the introduction, the primordial matter power spectrum can be seen as a largely model-independent parametrisation of the underlying theory. The goal of this section is to go from the power spectrum to parameters of specific cosmological models. This last step in the analysis can be seen as adding a layer to the Bayesian hierarchical model (see figure \ref{fig:BHM_cosmology}): $\boldsymbol{\upomega}$ is a vector of cosmological parameters which generates the primordial power spectrum coefficients $\boldsymbol{\uptheta}$. This generative process is usually deterministic, i.e.
\begin{equation}
\p(\boldsymbol{\uptheta}|\boldsymbol{\upomega}) = \updelta_\mathrm{D}\left( \boldsymbol{\uptheta} - \mathpzc{T}(\boldsymbol{\upomega}) \right),
\end{equation}
where $\mathpzc{T}$ is a deterministic function of cosmological parameters, typically a Boltzmann solver or a fitting function. Given this assumption, we have $\p(\boldsymbol{\Phi}|\boldsymbol{\upomega}) = \p(\boldsymbol{\Phi}|\mathpzc{T}(\boldsymbol{\upomega}))$, and the inference of cosmological parameters gives
\begin{equation}
\p(\boldsymbol{\upomega}|\boldsymbol{\Phi})_{|\boldsymbol{\Phi}=\boldsymbol{\Phi}_\mathrm{O}} = \widehat{L}_{\boldsymbol{\upomega}}^N(\boldsymbol{\upomega}) \frac{\p(\boldsymbol{\upomega})}{Z_{\boldsymbol{\Phi}}} ,
\label{eq:cosmological_inference_problem}
\end{equation}
where the likelihood for cosmological parameters is defined as
\begin{eqnarray}
\widehat{L}_{\boldsymbol{\upomega}}^N(\boldsymbol{\upomega}) & \equiv & \p(\boldsymbol{\Phi}|\mathpzc{T}(\boldsymbol{\upomega}))_{|\boldsymbol{\Phi}=\boldsymbol{\Phi}_\mathrm{O}} \nonumber\\
& = & \widehat{L}^N(\mathpzc{T}(\boldsymbol{\upomega})) .
\end{eqnarray}
Noting $\widehat{L}_{\boldsymbol{\upomega}}^N(\boldsymbol{\upomega}) \equiv\exp\left[ \hat{\ell}_{\boldsymbol{\upomega}}^N(\boldsymbol{\upomega}) \right]$ and using the linearised data model, we have (see equation \eqref{eq:linearised_effective_likelihood})
\begin{equation}
\begin{split}
-2 \hat{\ell}_{\boldsymbol{\upomega}}^N(\boldsymbol{\upomega}) \approx & \log\left| 2\pi\textbf{C}_0 \right| \\
& + \left[\boldsymbol{\Phi}_\mathrm{O} - \textbf{f}(\mathpzc{T}(\boldsymbol{\upomega}))\right]^\intercal \textbf{C}_0^{-1} \left[\boldsymbol{\Phi}_\mathrm{O} - \textbf{f}(\mathpzc{T}(\boldsymbol{\upomega}))\right].
\end{split}
\label{eq:likelihood_cosmological_parameters}
\end{equation}

Going from the primordial power spectrum to cosmological parameters is conceptually easy, but often hard in practice, when one only has samples of the power spectrum likelihood (or posterior). In this case, the likelihood to be used for cosmological parameter inference is naively represented by a sum of Dirac delta distributions. As it usually lives in a parameter space containing hundreds to thousands of dimensions, the number of samples is always a limiting factor. For this reason, techniques that effectively broaden the obtained samples, such as Blackwell-Rao estimators \citep{Wandelt2004} or kernel density estimates have been introduced. We note that our ABC technique does not suffer from this complication: the likelihood for cosmological parameters (equation \eqref{eq:likelihood_cosmological_parameters}) is a Gaussian centered on the data $\boldsymbol{\Phi}_\mathrm{O}$ with a fixed covariance matrix $\textbf{C}_0$. Furthermore, getting predictions for $\boldsymbol{\Phi}$ amounts to evaluating $\textbf{f}(\mathpzc{T}(\boldsymbol{\upomega}))$, which does not require additional black-box evaluations once $\textbf{f}$ is known. The Bayesian problem of inferring cosmological parameters $\boldsymbol{\upomega}$ (equation \eqref{eq:cosmological_inference_problem}) can therefore easily be solved by standard Markov Chain Monte Carlo techniques.

\section{Data model}
\label{sec:Data model}

In this section, we describe the generation of mock observations used to test the performance of our method. We emphasise that the statistical method presented in section \ref{sec:Method} is applicable to any black-box model, which can feature arbitrarily complex processes. Therefore, the details of the simulator presented in this section are of no relevance to the performance of the statistical method. It is only used in section \ref{sec:Results} as a showcase, to highlight the performance of our method to handle a complex simulator.

\subsection{Primordial power spectrum parametrisation}
\label{ssec:Primordial power spectrum parametrisation}

Throughout this paper, we work with a cubic equidistant grid with comoving side length of $1~\mathrm{Gpc}/h$ and $256^3$ voxels, spanning scales between $k_{s,\mathrm{min}} = 6.28 \times 10^{-3}$~$h$/Mpc and $k_{s,\mathrm{max}} = 1.4$~$h$/Mpc. We use $S=100$ support wavenumbers $k_s$. The first eight are fixed to the values required by the Fourier grid given our setup. The remaining support wavenumbers are logarithmically spaced up to $k_\mathrm{max}$. Any vector $\boldsymbol{\uptheta}$ in parameter space is defined at the scales of these support wavenumbers.

Between two consecutive support wavenumbers, we interpolate power spectra $P(k)$ using a one-dimensional spline fit, using $n=5$ as the degree of the smoothing spline. We checked that this setup yields vanishing differences in the representation of cosmological power spectra, at all wavenumbers of the Fourier grid used in this work.

\subsection{Galaxy surveys}
\label{ssec:Galaxy surveys}

The data model used in this work is a non-linear process meant to approximate the large variety of physical and observational phenomena at play in galaxy surveys. To do so, it uses the \textsc{Simbelmyn\"e} cosmological code, an end-to-end generative process for galaxy survey data given a specified primordial power spectrum $P(k)$. The flat $\Lambda$CDM model is assumed, and fiducial cosmological parameters used are the Planck 2015 values \citep[][table 4, last column]{PlanckCollaboration2015}, given in table \ref{tab:cosmotable}. These are used whenever the distance-redshift relation is needed, as well as for the gravitational evolution.\footnote{In principle, our approach would require treating cosmological parameters as nuisance parameters in the inference of the primordial power spectrum, and to marginalise over them. In this work, for simplicity, we keep them fixed to the values used to predict the expansion point.}
\begin{table}[h]
\centering
    \begin{tabular}{c c c c c c}
  \hline 
    $h$ & $\Omega_\mathrm{b}$&  $\Omega_\mathrm{m}$ & $n_\mathrm{S}$& $\sigma_8$ \\ \hline
	0.6674 & 0.0486  & 0.3089 & 0.9667  & 0.8159 \\ \hline 
    \end{tabular}
\caption{The cosmological parameters used in this work.\label{tab:cosmotable}} 
\end{table}

\begin{figure*}
\begin{center}
\includegraphics[width=\textwidth]{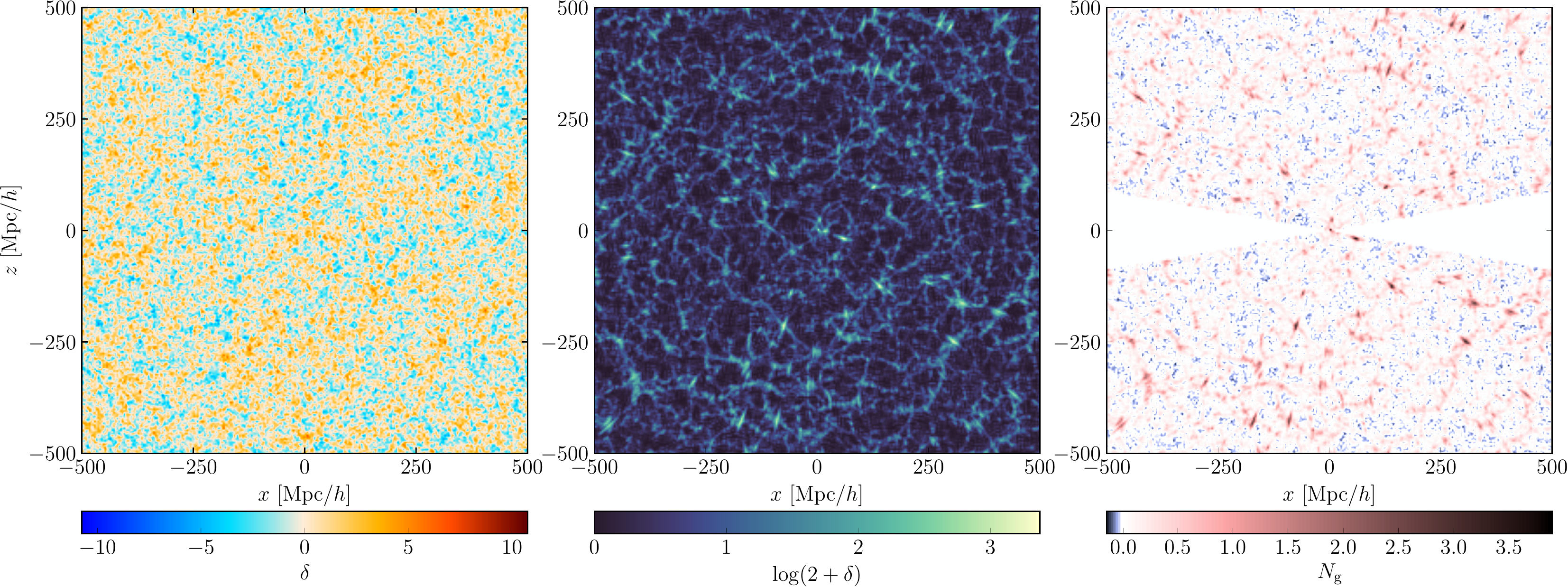}
\caption{Slices through one realisation of the initial density field (left) scaled to redshift zero with the linear growth factor, the corresponding evolved density field in redshift space (middle), and the field of observed galaxy number counts (right), according to the data model used in this work. The fields are defined on a grid of $256^3$ cells covering a total volume of $(1~\mathrm{Gpc}/h)^3$.\label{fig:slices}}
\end{center}
\end{figure*}

\textsc{Simbelmyn\"e} first generates a realisation of the initial density contrast via the convolution approach \citep[see e.g.][]{Peacock1985}. Specifically, the code generates a white noise field $\textbf{w}$, such that in each cell $x$ the value $w_x$ is drawn from the zero-mean unit-variance Gaussian distribution. The white noise field is multiplied by the square root of the desired cosmological power spectrum in Fourier space to give
\begin{equation}
\delta_k^\mathrm{i} \equiv \sqrt{P(k)} \, w_k .
\end{equation}
Transformation back to configuration space yields a initial density contrast field $\boldsymbol{\updelta}^\mathrm{i}$. One realisation of such an initial density field is shown in the left panel of figure \ref{fig:slices}.

Generated initial density fields then act as inputs to numerical structure formation simulations. The initial grid of $256^3$ voxels is populated by $512^3$ dark matter particles placed on a regular lattice. These particles are evolved to the redshift of $z=19$ via second order Lagrangian Perturbation theory (2LPT) \citep[see e.g.][]{Moutarde1991,Bouchet1995,Bouchet1996}, then with an efficient implementation of \textsc{cola} (COmoving Lagrangian Acceleration, \citealp{Tassev2013}) from $z=19$ to $z=0$. A particle-mesh grid of $1024^3$ voxels and 20 timesteps linearly-spaced in the scale factor are used for the evolution with \textsc{cola}. We checked that, at the scales of interest, this setup yields negligible difference in the representation of final density fields with respect to the prediction of the fully non-linear code \textsc{gadget-2} \citep{Springel2005}. In particular, final density fields contain the additional power expected from non-linear structure formation, at 1\% precision up to $k = 0.5$ $h/\mathrm{Mpc}$ and 5\% up to $k = 1$ $h/\mathrm{Mpc}$. We place the observer at the centre of the box. The maximal distance to the observer is $866$~Mpc/$h$, which we consider sufficiently small to neglect light-cone effects. Therefore, in this paper, we only use the final snapshot of our simulations at redshift zero, and ignore the evolution of matter within the survey volume.

Using their final peculiar velocities with respect to the observer $v_r$, dark matter particles are placed in redshift space according to the non-linear mapping 
\begin{equation}
1+z_\mathrm{obs} = (1+z_\mathrm{cosmo})(1+z_\mathrm{pec}), \enspace \mathrm{with} \enspace  z_\mathrm{pec} \equiv -\frac{v_r}{\mathrm{c}},
\end{equation}
where $z_\mathrm{cosmo}$ is the true cosmological redshift, $z_\mathrm{obs}$ is the ``observed'' redshift and $\mathrm{c}$ is the speed of light. Note that we do not work in the plane-parallel approximation. The particles are then binned to a $256^3$-voxels grid with the cloud-in-cell scheme \citep{Hockney1981} to give the final density contrast field $\boldsymbol{\updelta}^\mathrm{f}$. One realisation of the final redshift-space density is shown in the middle panel of figure \ref{fig:slices}.

The galaxy density $\boldsymbol{\uprho}^\mathrm{g}$ is predicted using a linear bias model, used in various previous studies \citep[e.g.][]{Verde2002,Ross2015}, such that in any cell $x$,
\begin{equation}
\rho_x^\mathrm{g} = \bar{N} \, (1 + b \, \delta_x^\mathrm{f}) .
\end{equation}
In this work, we use $b=1.2$ and $\bar{N}=0.119$ (corresponding to an observed galaxy number density $\bar{n} = 2 \times 10^{-3}~(h/\mathrm{Mpc})^3$, achievable for instance with the Euclid spectroscopic survey, see e.g. \citealp{Majerotto2012}). For the sake of simplicity, $b$ and $\bar{N}$ are fixed, but they could be straightforwardly treated as additional nuisance parameters and marginalised over.

\begin{figure}
\begin{center}
\includegraphics[width=\columnwidth]{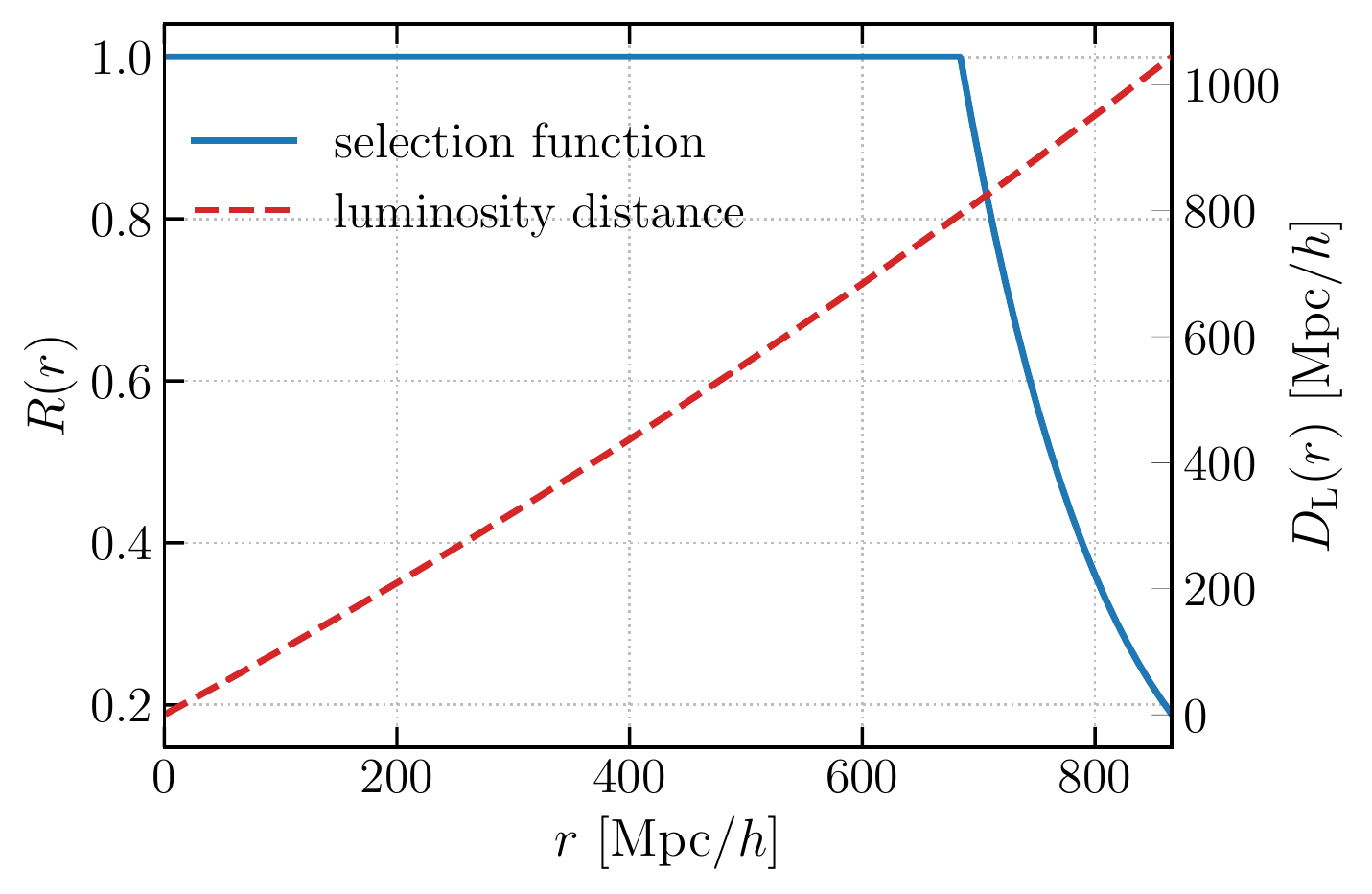}
\caption{Radial selection function $R(r)$ (solid blue line) used in our synthetic survey as a function of the comoving distance to the observer $r$. The model used is a Schechter luminosity function with parameters $\alpha=-1.05$ and $M_\ast = -20.44$, a limiting apparent magnitude of $m = 18.5$ and absolute magnitude cuts $-25 \leq M \leq -21$. The luminosity distance $D_\mathrm{L}$ (dashed red line, right vertical axis) is also shown. The survey is complete up to $D_\mathrm{L}=794$~Mpc/$h$, corresponding to $r=684$~Mpc/$h$.\label{fig:selection_function}}
\end{center}
\end{figure}

The last step corresponds to a virtual observation of the galaxy field, accounting for observational effects expected in actual surveys. To do so, we use the three-dimensional survey response operator (or window) $\textbf{W}$, consisting of the product of the radial selection function $R(r)$ and the angular survey mask and completeness function $C(\hat{\textbf{n}})$ for any line-of-sight $\hat{\textbf{n}}$, i.e. $\textbf{W}(\hat{\textbf{n}},r) \equiv R(r) \, C(\hat{\textbf{n}})$. This operator accounts for the fact that we are only looking at certain parts of the sky and that we have different detection probabilities of galaxies depending on their distance. We obtained a simple model for $\textbf{W}$ on a grid matching our simulations ($256^3$ voxels covering a volume of $(1~\mathrm{Gpc}/h)^3$) as follows. For the angular completeness, we mask the galactic plane by setting $C(\hat{\textbf{n}})$ to $0$ for galactic latitudes $-10\degree \leq b \leq 10\degree$ and $1$ otherwise, the system of coordinates being defined such that the observer is at the origin and the plane of equation $z=0$ is the galactic plane. For the radial selection function $R(r)$, we use a Schechter luminosity function \citep{Schechter1976} with previously-published parameters for the $r$-band: $\alpha=-1.05$ and $M_\ast = -20.44$ \citep{Blanton2003b}, a limiting apparent magnitude of $m = 18.5$ and absolute magnitude cuts $-25 \leq M \leq -21$. This choice makes our synthetic survey complete up to a luminosity distance of $D_\mathrm{L}=794$~Mpc/$h$, corresponding to $r=684$~Mpc/$h$ (see figure \ref{fig:selection_function}).

We emulate the survey via galaxy number counts $N_x^{\mathrm{g}}$ in each cell $x$ of the box. In order to account for instrumental noise, our model is a non-uniform Gaussian process, meaning that in every cell $N_x^{\mathrm{g}}$ is drawn from a Gaussian distribution with mean $\mu^\mathrm{g}_x$ and standard deviation $\sigma_x$,
\begin{equation}
N_x^{\mathrm{g}} \curvearrowleft \mathpzc{G}(\mu^\mathrm{g}_x|\sigma_x)
\end{equation}
The mean $\mu^\mathrm{g}_x \equiv W_x \rho^\mathrm{g}_x = W_x \bar{N} (1+ b\, \delta^\mathrm{f}_x)$ characterises the expected number of galaxies. It accounts for all physical effects via $\delta^\mathrm{f}_x$ and for the survey response operator $W_x$. The standard deviation is defined as $\sigma_x \equiv \sigma \sqrt{W_x\bar{N}}$, where we set the overall noise level $\sigma \equiv 10^{-1}$. A slice through one realisation of the galaxy number count field $\textbf{N}^{\mathrm{g}}$ is shown in the right panel of figure \ref{fig:slices}.

\subsection{Summary statistic}
\label{ssec:Summary statistic}

As is standard practice in cosmological data analysis, we do not work at the level of the entire galaxy number count map $\textbf{N}^{\mathrm{g}}$, but use a compression $\mathpzc{C}$ of the map to well-chosen statistical summaries. In this work, we limit ourselves to an estimator of the final power spectrum of the survey. It should be remarked however, that the method can use any other statistical summary and even combinations of those.

We obtain the binned data power spectrum by taking the squared modulus of the Fourier transform of $\textbf{N}^\mathrm{g}$, summing the contributions over all the Fourier cells within wavenumber shell $k_r$, and normalizing:
\begin{equation}
P^\mathrm{f}(k_r) \equiv C \times \sum_{|\textbf{k}| \in k_r} \frac{\left| N_k^\mathrm{g} \right|^2}{N_{k_r}-2} .
\end{equation}
The overall constant factor $C = 1000^3/256^2~(\mathrm{Mpc}/h)^3$ arises from our Fourier transform convention; $N_{k_r}$ represents the number of modes within the shell $k_r$; and the factor $-2$ arises from the assumption that the data power spectrum is inverse-$\Gamma$ distributed with shape parameter $N_{k_r}/2$ and scale parameter $C \times \left| N^\mathrm{g}_k \right|^2/2$ \citep[see][]{Jasche2010b}.

As the galaxy data model relies on a full cosmological simulation, the predictions for the final power spectrum remain reasonable even at scales that have experienced substantial non-linearity. In spite of the approximations made, we trust our data model at the percent level up to $k = 0.5$~$h$/Mpc. For this reason, we use $P=43$ $k_r$-bins in the range $[0.02,0.5]$ $h$/Mpc, ensuring that each bin contains at least $100$ modes. These bins are logarithmically spaced for $k_r \geq 0.04$ $h$/Mpc. For convenience, we normalise the output of the black-box using the expansion point, so that $\boldsymbol{\Phi}$ is the $P$-dimensional vector of components $A \times P^\mathrm{f}(k_r)/P_0(k_r)$ with $A = 50$. The estimation of $P^\mathrm{f}(k)$ is completely deterministic once the galaxy number counts are given. Therefore, we now have a complete model to generate artificial realisations of $\boldsymbol{\Phi}$ for a given primordial power spectrum $P(k)$ and specific realisations of initial phases and noise.

Due to the breakdown of models based on perturbation theory in the non-linear regime and to the difficulties in incorporating the impact of small-scale observational processes, state-of-the-art large-scale structure analyses are typically limited to $k_\mathrm{max} \lesssim 0.3$ $h$/Mpc \citep[e.g.][]{Ross2015}. Pushing the analysis to $k_\mathrm{max} = 0.5$~$h$/Mpc represents an increase by a factor of $\sim 5$ in the number of modes used (scaling as $k_\mathrm{max}^3$), which is expected to yield substantial improvements in the inference results.

\subsection{Idealised data model}
\label{ssec:Idealised data model}

For testing purposes, we also define an idealised data model corresponding to a Gaussian random field. More specifically, using exactly the same setup as before, the black-box here simply consists of producing the initial density field $\boldsymbol{\updelta}_\mathrm{i}$ (see section \ref{ssec:Galaxy surveys}), scaling it to redshift zero using the linear growth factor, and measuring its normalised power spectrum $\boldsymbol{\Phi}$ (see section \ref{ssec:Summary statistic}). The use of the two different black-boxes (Gaussian random field and realistic mock survey) within our method will quantify the effect of non-linear gravity, redshift-space distortions, and survey complications on primordial power spectrum inference, in particular the detectability of BAOs.

\section{Results}
\label{sec:Results}

This section describes the results obtained by applying the statistical method proposed in section \ref{sec:Method} in conjunction with the data-generating process described in section \ref{sec:Data model} to an artificial galaxy survey, itself generated using the same process.

We use for $P_0(k)$ the ``wiggle-less'' BBKS power spectrum \citep{Bardeen1986} under Planck 2015 cosmology (see table \ref{tab:cosmotable}). Unknown ground truth cosmological parameters $\boldsymbol{\upomega}_\mathrm{gt}$ are drawn from the (marginalised, Gaussian) Planck priors:
\begin{equation}
\begin{pmatrix}
h \\
\Omega_\mathrm{b} \\
\Omega_\mathrm{m} \\
n_\mathrm{S} \\
\sigma_8
\end{pmatrix} \curvearrowleft \mathpzc{G} \left[
\begin{pmatrix}
0.6774 \\
0.04860 \\
0.3089 \\
0.9667 \\
0.8159 \\
\end{pmatrix},
\mathrm{diag}
\begin{pmatrix}
0.0046^2 \\
0.00030^2 \\
0.0062^2 \\
0.0040^2 \\
0.0086^2
\end{pmatrix} \right].
\label{eq:Planck_priors}
\end{equation}
The ``wiggly'' ground truth power spectrum $P_{\textrm{gt}}(k)$ is generated with the \citet{Eisenstein1998} (EH) fitting function, using these cosmological parameters. It is used to simulate observed data $\boldsymbol{\Phi}_\mathrm{O}$, with unknown nuisance parameters (phase realisation and instrumental noise). For later use, the fiducial ``wiggly'' power spectrum $P_\mathrm{fid}(k)$ is also generated with the EH prescription, using Planck cosmology. The target parameters $(\boldsymbol{\uptheta})_s \equiv P(k_s)/P_0(k_s)$ are the values of the wiggle function at the $S=100$ support wavenumbers defined in section \ref{ssec:Primordial power spectrum parametrisation}. We note $\boldsymbol{\uptheta}_\mathrm{gt}$ and $\boldsymbol{\uptheta}_\mathrm{fid}$ the vectors of component $P_\mathrm{gt}(k_s)/P_0(k_s)$ and $P_\mathrm{fid}(k_s)/P_0(k_s)$, respectively.

\subsection{Diagnostics of the black-box}
\label{ssec:Diagnostics of the black-box}

\begin{figure*}
\begin{center}
\includegraphics[width=\textwidth]{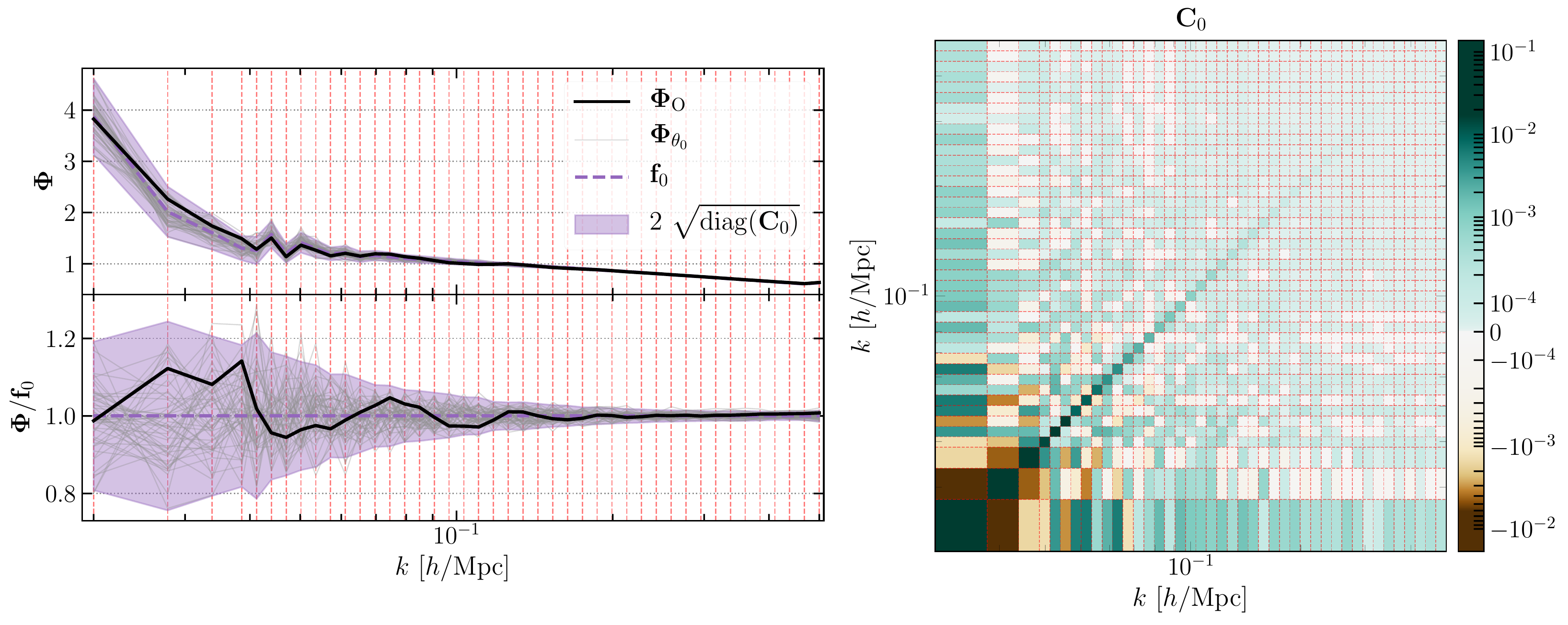}
\caption{Diagnostics of the black-box at the expansion point $\boldsymbol{\Phi}_0$. The left panel shows individual mock observations $\boldsymbol{\Phi}_{\boldsymbol{\uptheta}_0}$ (grey lines), the observed data $\boldsymbol{\Phi}_\mathrm{O}$ (solid black line), and the average black-box $\textbf{f}_0$ (dashed purple line). The shaded region corresponds to two standard deviations. The right panel shows the covariance matrix of the summaries at the expansion point, $\textbf{C}_0$. The dashed red lines correspond to the positions of the bins at which summaries are measured in data space.\label{fig:blackbox_expansion}}
\end{center}
\end{figure*}

\begin{figure*}
\begin{center}
\includegraphics[width=\textwidth]{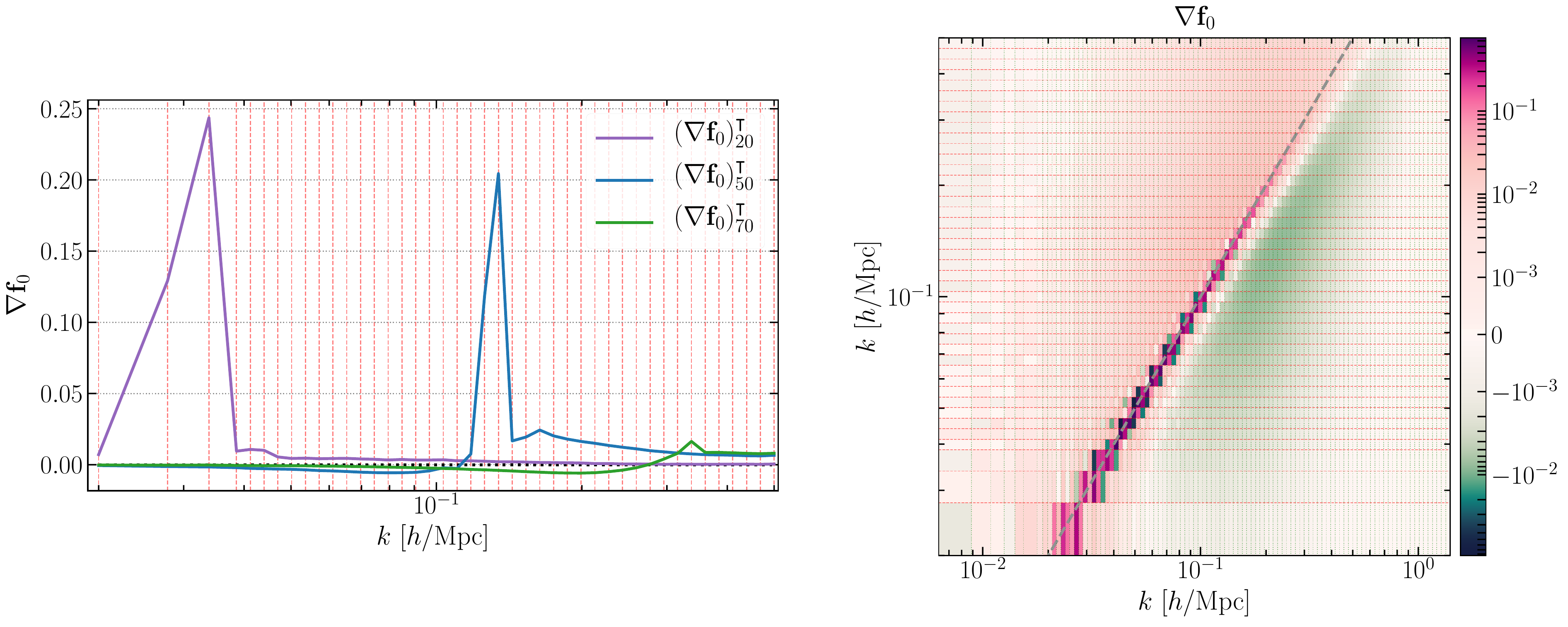}
\caption{Gradient of the black-box, measured via finite differencing. The right panel shows individual columns $(\nabla \textbf{f}_0)^\intercal_s$ for $s =20, 50$, and $70$, corresponding to support wavenumbers $k_s = 0.0364, 0.1484$, and $0.3780$~$h$/Mpc, respectively. The right panel shows the gradient matrix $\nabla \textbf{f}_0$. The dashed grey line corresponds to the identity function; the dashed red lines correspond to the positions of the bins at which summaries are measured in data space; and the dotted green lines correspond to the support wavenumbers in parameter space.\label{fig:blackbox_gradient}}
\end{center}
\end{figure*}

We created an ensemble of $N_0=150$ mock realisations at the expansion point $\boldsymbol{\uptheta}_0 = \boldsymbol{1}_{\mathbb{R}^S}$ using different nuisance parameters. These are used to compute the average black-box at the expansion point, $\textbf{f}_0 \equiv \boldsymbol{\hat{\Phi}}_{\boldsymbol{\uptheta}_0}$, and the covariance matrix $\textbf{C}_0 \equiv \boldsymbol{\hat{\Sigma}}'_{\boldsymbol{\uptheta}_0}$, using their definitions (equations \eqref{eq:estimated_mean}, \eqref{eq:covariance_effective_likelihood}, and \eqref{eq:estimated_covariance}). The results are shown in figure \ref{fig:blackbox_expansion}. There, the left panel shows individual realisations $\boldsymbol{\Phi}_{\boldsymbol{\uptheta}_0}$ and the observed data vector $\boldsymbol{\Phi}_\mathrm{O}$. The average black-box $\textbf{f}_0$ is also plotted, with a credible region corresponding to two standard deviations (i.e. $2 \sqrt{\mathrm{diag}(\textbf{C}_0})$). The full estimated covariance matrix $\textbf{C}_0$ is shown in the right panel. As expected, the measured variance is larger on large scales due to cosmic variance, with some anti-correlations between pairs of bins. The effect of the mask, which increases power at the largest scales found in the simulation box, is also clearly visible.

Using a step size of $h=10^{-2}$ and an ensemble of $N_s = 100$ mock realisations at each of the expansion points $\boldsymbol{\uptheta}_s$, we measured the gradients of the black-box $(\nabla \textbf{f}_0)^\intercal_s$ along all directions of parameter space (see equation \eqref{eq:gradient_finite_differencing}). The nuisance parameters (phase realisation and noise) are kept at fixed values (the ones corresponding to the first $N_s$ realisations generated at the expansion point) for this calculation. The results are shown in figure \ref{fig:blackbox_gradient}, where the left panel shows $(\nabla \textbf{f}_0)^\intercal_s$ for individual values of $s$ and the right panel shows the full rectangular matrix $\nabla \textbf{f}_0$. Some interesting phenomena can be observed. At large scales, (see e.g. for $k_{20}=0.0364$~$h$/Mpc) exciting one initial mode ony triggers an answer in the bins closest to this scale; the gradient therefore resembles a multiple of the identity function. This is the result expected from linear perturbation theory. However, at small scales, the non-linear simulator couples modes. This implies that the response is smaller in amplitude but distributed over a much larger ranges of scales (see e.g. for $k_{70} = 0.3780$~$h$/Mpc). In the non-linear regime, the gradient is typically negative at large scales, crosses zero slightly before the excited scale $k_s$, then becomes positive at smaller scales.

As discussed in section \ref{ssec:Linearisation of black-box models}, the linearised black-box $\textbf{f}(\boldsymbol{\uptheta})$ is fully characterised by $\textbf{f}_0$, $\textbf{C}_0$ and $\nabla \textbf{f}_0$. In this work, we used a total of $N_0 + N_s \times S = 10,150$ simulations to get very precise estimates of $\textbf{C}_0$ and $\nabla \textbf{f}_0$, although using fewer would have been possible.

\subsection{The prior and its optimisation}
\label{ssec:The prior and its optimisation}

\begin{figure}
\begin{center}
\includegraphics[width=\columnwidth]{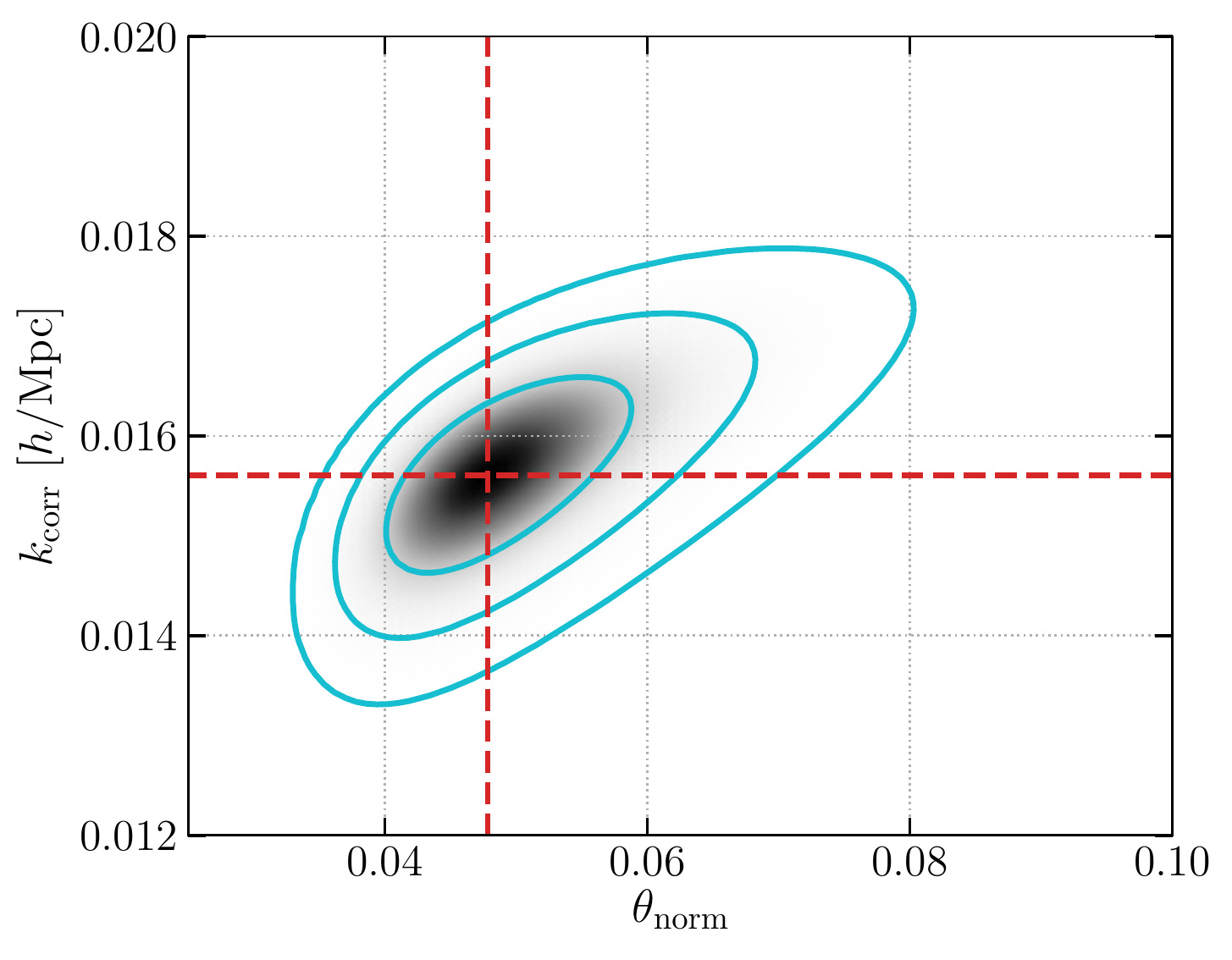}
\caption{Posterior surface for the prior hyperparameters $k_\mathrm{corr}$ and $\theta_\mathrm{norm}$. The $1\sigma$, $2\sigma$ and $3\sigma$ credible contours are shown as solid blue lines. The dashed red lines mark the maximum \textit{a posteriori} values.\label{fig:hyperparameters}}
\end{center}
\end{figure}

\begin{figure*}
\includegraphics[width=\textwidth]{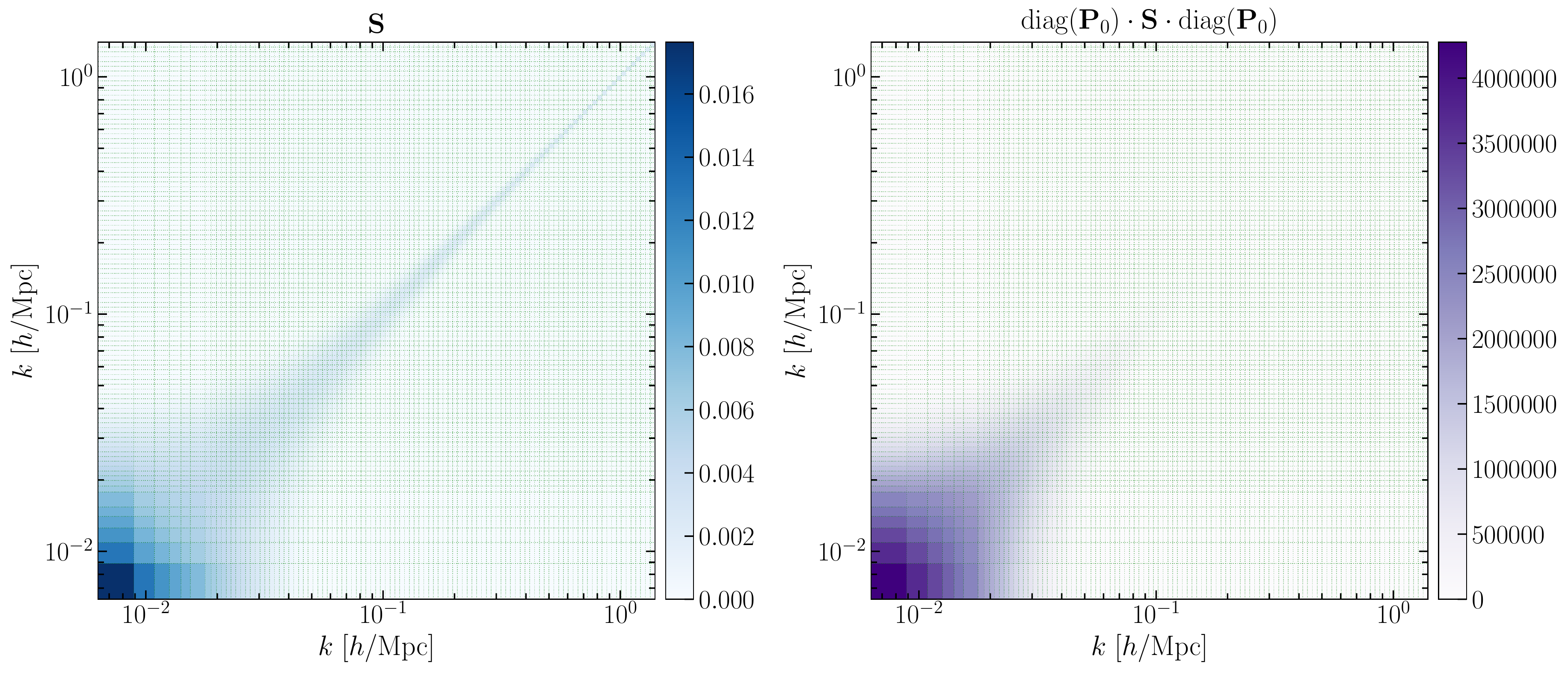}
\caption{Prior covariance matrix $\textbf{S}$ for the wiggle function $\boldsymbol{\uptheta}$ (left panel). The corresponding prior covariance matrix for power spectrum amplitudes, $\mathrm{diag}(\textbf{P}_0) \cdot \textbf{S} \cdot \mathrm{diag}(\textbf{P}_0)$, is shown in the right panel. The dotted green lines correspond to the support wavenumbers in parameter space.\label{fig:prior_covariance}}
\end{figure*}

As discussed in section \ref{ssec:The power spectrum prior distribution}, we choose a Gaussian prior centered on the expansion point $\boldsymbol{\uptheta}_0$ with a covariance matrix $\textbf{S}$ given by equation \eqref{eq:prior_covariance} and characterised by three hyperparameters $\{k_\mathrm{corr}, \alpha_\mathrm{cv}, \theta_\mathrm{norm}\}$.

Following the method presented in section \ref{ssec:Optimisation of the prior hyperparameters}, we found optimal values for the prior hyperparameters. The strength of cosmic variance within our simulation volume shall satisfy $\alpha_\mathrm{cv} = \sqrt{k^3/N_k}$ at all scales $k$, where $N_k$ is the number of modes. In our Fourier grid (described in section \ref{ssec:Primordial power spectrum parametrisation}), we measured up to the Nyquist frequency $\alpha_\mathrm{cv} = 8.848 \times 10^{-4}$, value that we adopt.

Assuming that the target function $\boldsymbol{\uptheta}$ follows the functional shape of the fiducial wiggle function $\boldsymbol{\uptheta}_\mathrm{fid}$ calculated with Planck cosmology, the likelihood for $k_\mathrm{corr}$ and $\theta_\mathrm{norm}$ is given by equation \eqref{eq:likelihood_hyperparameters}. We further assume broad, uncorrelated Gaussian hyperpriors on $k_\mathrm{corr}$ and $\theta_\mathrm{norm}$: $k_\mathrm{corr} \sim \mathpzc{G}(0.020, 0.015^2)$ [$h$/Mpc] and $\theta_\mathrm{norm} \sim \mathpzc{G}(0.2, 0.3^2)$. The posterior surface is plotted in figure \ref{fig:hyperparameters}. We found the maximum \textit{a posteriori} values using the popular optimiser L-BFGS \citep{L-BFGS}. In our run, these are $k_\mathrm{corr} = 0.0156$~$h$/Mpc and $\theta_\mathrm{norm} = 0.0478$. The resulting prior covariance matrix is shown in figure \ref{fig:prior_covariance}. For the idealised data model (see section \ref{ssec:Idealised data model}), the same procedure is applied and the optimal parameters are found to be $k_\mathrm{corr} = 0.0158$~$h$/Mpc and $\theta_\mathrm{norm} = 0.0535$.

\subsection{The effective posterior}
\label{ssec:The effective posterior}

\begin{figure*}
\begin{center}
\includegraphics[width=0.9\textwidth]{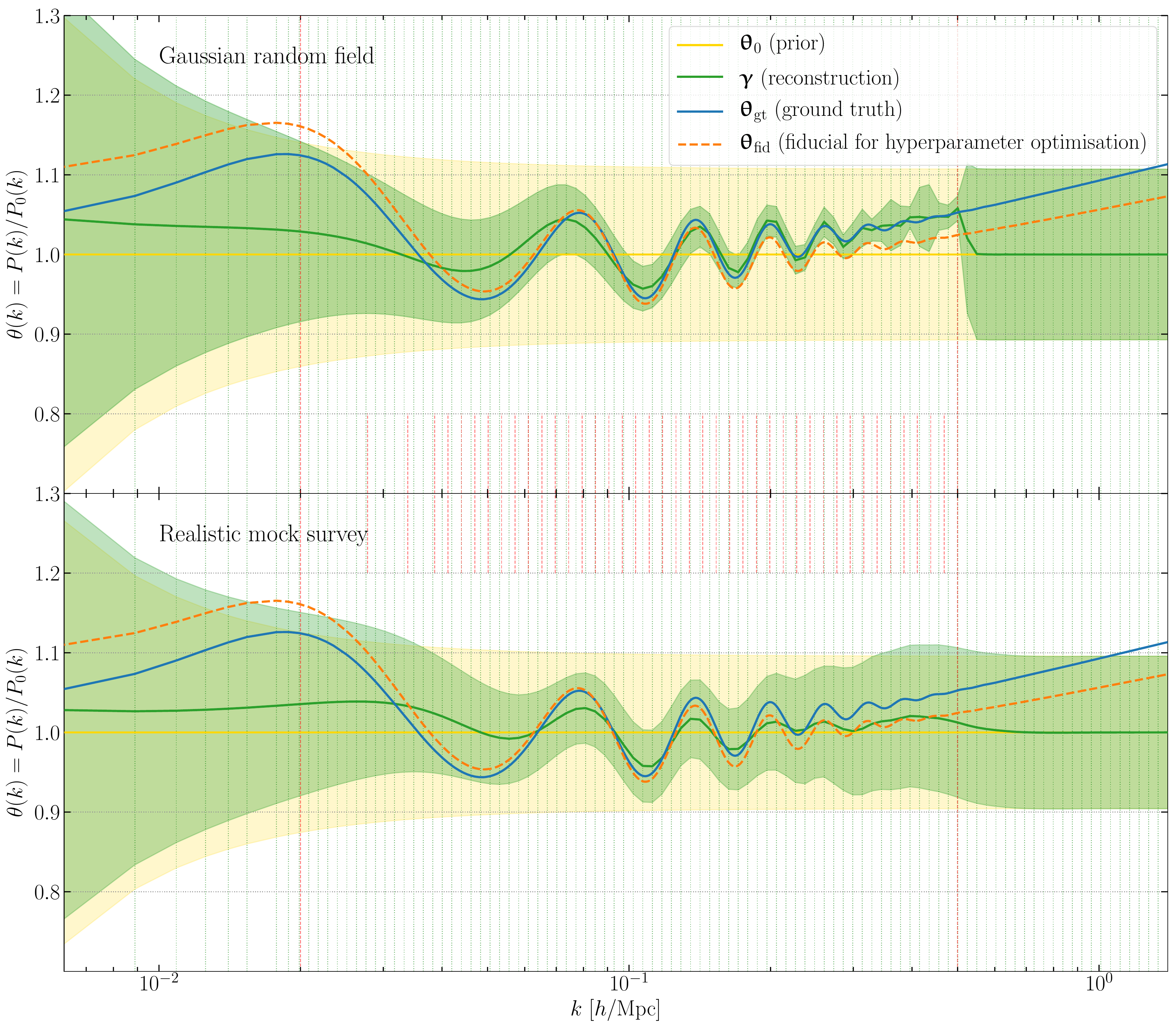}
\caption{Inference of the wiggle function $\theta(k) \equiv P(k)/P_0(k)$ as a function of wavenumber $k$, using as data model a Gaussian random field (top panel) or a realistic mock survey (bottom panel). The prior mean $\boldsymbol{\uptheta}_0$ and the effective posterior mean $\boldsymbol{\upgamma}$ are represented as solid yellow and green lines, respectively, with their $2\sigma$ credible intervals (for the prior, $2\sigma = 2 \,\theta_\mathrm{norm}(1+\alpha_\mathrm{cv}/k^{3/2})$). For comparison, the ground truth $\boldsymbol{\uptheta}_\mathrm{gt}$ and the fiducial ``wiggly'' function $\boldsymbol{\uptheta}_\mathrm{fid}$ used to optimise the prior hyperparameters are plotted as solid blue and dashed orange line, respectively. The dashed red lines correspond to the positions of the bins at which summaries are measured in data space; and the dotted green lines correspond to the support wavenumbers in parameter space. In the realistic case, in spite of survey complications which limit the information captured, the signature of BAOs is well reconstructed up to $k \approx 0.3$ $h$/Mpc, with 5 inferred acoustic peaks, result which could be improved using more volume. In the absence of informative data, the power spectrum reconstruction is driven towards the prior mean, but this effect does not affect cosmological parameter inference (see section \ref{ssec:The effective posterior} for details).\label{fig:reconstruction}}
\end{center}
\end{figure*}

\begin{figure*}
\begin{center}
\includegraphics[width=\textwidth]{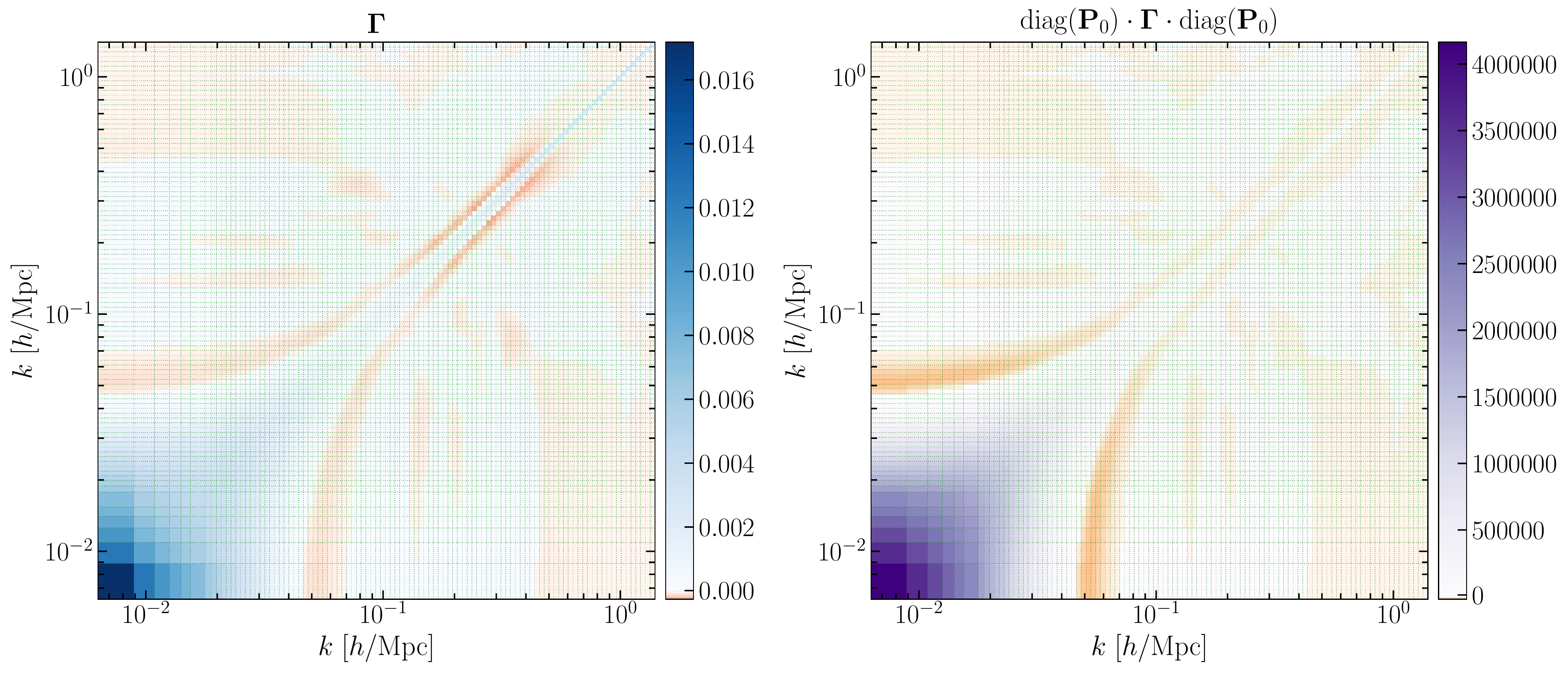}
\caption{Posterior covariance matrix $\boldsymbol{\Gamma}$ for the wiggle function $\boldsymbol{\uptheta}$ (left panel). The corresponding posterior covariance matrix for power spectrum amplitudes, $\mathrm{diag}(\textbf{P}_0) \cdot \boldsymbol{\Gamma} \cdot \mathrm{diag}(\textbf{P}_0)$, is shown in the right panel. The dotted green lines correspond to the support wavenumbers in parameter space. The correlation structure between different inferred parameters exhibits a complex behaviour (see also figure \ref{fig:prior_covariance} for comparison).\label{fig:posterior_covariance}}
\end{center}
\end{figure*}

Using the effective likelihood, characterised by the observed data $\boldsymbol{\Phi}_\mathrm{O}$ and the linearised black-box described in section \ref{ssec:Diagnostics of the black-box}, as well as the optimised prior discussed in section \ref{ssec:The prior and its optimisation}, we obtained the effective posterior on $\boldsymbol{\uptheta}$. It is a Gaussian, with mean and covariance matrix given by the ``filter equations'' \eqref{eq:filter_mean} and \eqref{eq:filter_var}.

Figure \ref{fig:reconstruction} shows the inferred primordial wiggle function $\boldsymbol{\upgamma}$ in comparison with the expansion point $\boldsymbol{\uptheta}_0$, the ground truth $\boldsymbol{\uptheta}_\mathrm{gt}$, and the fiducial function used to optimise prior hyperparameters, $\boldsymbol{\uptheta}_\mathrm{fid}$. $2\sigma$ credible regions are shown for the prior and the posterior (i.e. $2 \sqrt{\mathrm{diag}(\textbf{S})}$ and $2 \sqrt{\mathrm{diag}(\boldsymbol{\Gamma})}$, respectively). The top panel corresponds to the result obtained using the idealised data model (a Gaussian random field, see section \ref{ssec:Idealised data model}), and the bottom panel to the result obtained using the realistic mock survey data model. In both cases, the inference is unbiased since the ground truth always lies within the $2\sigma$ credible intervals of the reconstruction. As discussed in the introduction, this effective posterior can be seen as a largely model-independent parametrisation of the theory, containing all the available cosmological information under weak assumptions.

As can be read from the figure, the inferred vector contains the BAO wiggles, even far within the Silk damping tail \citep{Silk1968}. All visible oscillations are fully reconstructed in the idealised case, and in the realistic case, up to scales of $k \approx 0.3$~$h$/Mpc. In particular, 5 acoustic oscillations are unambiguously identified, which is competitive with the latest cosmic microwave background experiments and has been so far out of reach of galaxy surveys. Note that this result is obtained given a simulation volume of $(1~\mathrm{Gpc}/h)^3$ and that further improvements could be obtained with a larger volume, as will be probed by upcoming surveys. The inferred wiggle function shows higher uncertainty in regions of small and large wavenumbers. This is due to cosmic variance and noise, respectively. Cosmic variance reflects the limited number of modes that we have at the largest scales in our simulation volume. This effect limits the significance of the determination of the cosmological power spectrum at these scales. For this reason, at $k \lesssim 0.05$ $h$/Mpc the reconstruction is driven towards the prior mean, which is also the expansion point and the default answer in the absence of informative data. As expected, this effect is visible in the idealised as well as in the realistic case. On the other hand, noise (understood as the combined effect of the specific phase realisation of the data, non-linear gravity, redshift-space distortions and instrumental noise) acts on smaller scales ($0.2~h/\mathrm{Mpc} \lesssim k \lesssim 0.5~h/\mathrm{Mpc}$). Some of the primordial information is effectively destroyed at these scales -- or at least is not captured by the statistical summaries $\boldsymbol{\Phi}$. The reconstruction is therefore also driven towards the prior mean, and the uncertainty is increased, because the data are less informative than in the idealised case. As expected, we recover the prior at $k \gtrsim 0.5$ $h$/Mpc (in fact a little below, due to mode coupling), since the data $\boldsymbol{\Phi}$ do not contain measurements at these scales.

It is important to note that the prior $\p(\boldsymbol{\uptheta})$, used in this section to regularise the inference of the primordial power spectrum, does not appear in the inference of cosmological parameters in the next section (only the effective likelihood does). Thus, no bias is introduced in cosmological parameter inference when the amplitude of reconstructed BAO wiggles seems to undershoot the ground truth.

Since the proposed method is fully Bayesian, we do not simply obtain a point estimate, but a complete probability distribution, which provides a detailed quantification of uncertainties. Figure \ref{fig:posterior_covariance} shows the covariance matrix $\boldsymbol{\Gamma}$ of the Gaussian effective posterior; it can be compared with the prior covariance matrix $\textbf{S}$ shown in figure \ref{fig:prior_covariance}.

\subsection{Cosmological parameters}
\label{ssec:Cosmological parameters}

\begin{figure*}
\begin{center}
\includegraphics[width=\textwidth]{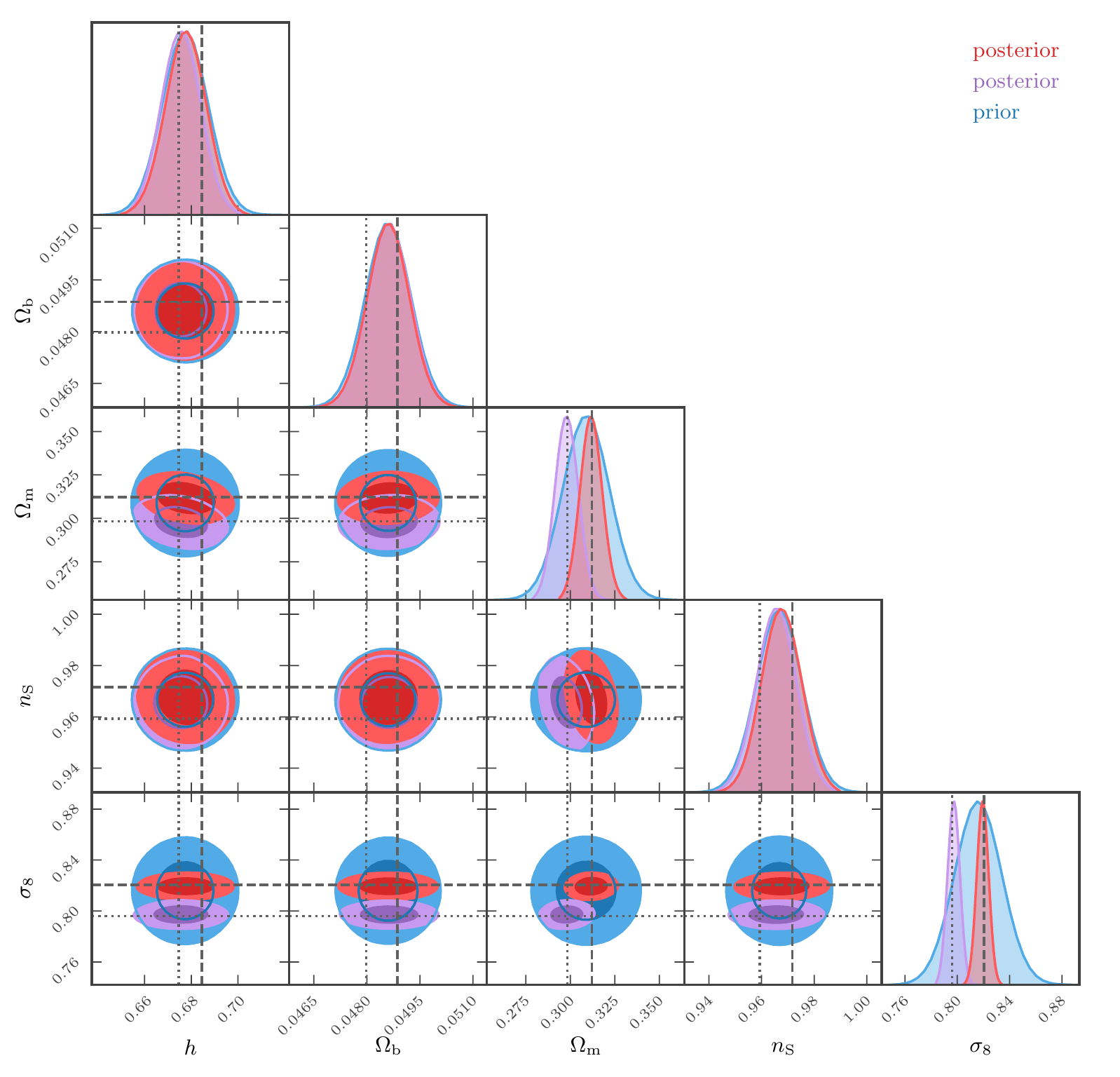}
\caption{Cosmological parameter inference using a linearised black-box model of galaxy surveys. The prior is shown in blue, and the effective posteriors for two different data realisations are shown in red and purple. The two different data realisations have different input cosmological parameters (shown as dashed and dotted lines), different phases and noise realisations. For all distributions, the $1\sigma$ and $2\sigma$ contours are shown.\label{fig:cosmology}}
\end{center}
\end{figure*}

The last step in the analysis is to infer parameters $\boldsymbol{\upomega}$ of specific cosmological models, given the observed data $\boldsymbol{\Phi}_\mathrm{O}$. In this section, we assume a flat $\Lambda$CDM model, characterised by $5$ parameters $\{h, \Omega_\mathrm{b}, \Omega_\mathrm{m}, n_\mathrm{S}, \sigma_8 \}$. For simplicity, our prescription to generate primordial power spectra given cosmological parameters is the EH fitting function. The generative process $\mathpzc{T}$ (from $\boldsymbol{\upomega}$ to $\boldsymbol{\uptheta}$) is therefore the EH fitting function divided by $P_0$, sampled at the support wavenumbers $k_s$. It could easily be generalised to extensions of the flat $\Lambda$CDM model and to include a Boltzmann solver. The linearised data model $\textbf{f}$ (from $\boldsymbol{\uptheta}$ to $\boldsymbol{\Phi}$) has already been characterised in section \ref{ssec:Diagnostics of the black-box}.

The effective likelihood $\widehat{L}_{\boldsymbol{\upomega}}^N(\boldsymbol{\upomega})$ for cosmological parameters is given by equation \eqref{eq:likelihood_cosmological_parameters}. Consistently with the expansion point used to linearise the black-box, we complement $\widehat{L}_{\boldsymbol{\upomega}}^N(\boldsymbol{\upomega})$ with a Gaussian prior $\p(\boldsymbol{\upomega})$ centred on Planck cosmological parameters, but with broader variance: the diagonal covariance matrix given in equation \eqref{eq:Planck_priors} is multiplied by a factor of $3$. We explored the effective posterior $\p(\boldsymbol{\upomega}|\boldsymbol{\Phi})_{|\boldsymbol{\Phi}=\boldsymbol{\Phi}_\mathrm{O}}$ (equation \eqref{eq:cosmological_inference_problem}) via MCMC (performed using the \textsc{emcee} code, \citealp{Foreman-Mackey2013}), ensuring sufficient convergence.

Results are shown in figure \ref{fig:cosmology}. Prior contours are shown in blue, and posterior contours are shown in red and purple for two different realisations of the data $\boldsymbol{\Phi}_\mathrm{O}$. The two data realisations have been generated using different ground truth cosmological parameters $\boldsymbol{\upomega}_\mathrm{gt}$ (shown as dashed and dotted lines, respectively), as well as different nuisance parameters (phase realisation and noise). The plot demonstrates that cosmological parameter inference is unbiased and robust to nuisances imprinted in the data.

\section{Discussion and conclusion}
\label{sec:Discussion and conclusion}

The biggest challenge in galaxy survey analyses arises from the requirement of non-linear data modelling. In this work, we described the development of a novel simulation-based Bayesian approach, {\selfi}, which can be used to infer the primordial matter power spectrum and cosmological parameters from galaxy surveys. The main results are the ``filter equations'' \eqref{eq:filter_mean} and \eqref{eq:filter_var}. They can be applied to get an effective posterior for any model where the mean and covariance of the data are estimated from arbitrarily complex forward models. Essentially everything is obtained from a simulator, which can be treated as a black-box, without necessity to include any knowledge of its internal mechanisms into the statistical analysis. 

We derived the ``filter equations'' under two assumptions: the availability of a black-box able to generate artificial data, and of strong prior constraints in parameter space, obtained from a previous experiment. We built an effective likelihood for this scenario and made its evaluation efficient by linearising the black-box around an expansion point. We devised a method to optimise the hyperparameters appearing in our power spectrum prior. Finally, we derived the cheap likelihood for parameters of specific cosmological models, to be used in our framework.

The approach presented in this paper relies on likelihood-free forward-modelling via ABC. It complements statistically exact, likelihood-based forward-modelling techniques. The principal differences are as follows.
\begin{itemize}
\item First, the numerical complexity of likelihood-based MCMC approaches typically requires to approximate complex data models to allow for fast execution speeds. In this work, we rather aimed at performing approximate inference, but with full-scale black-box models. This approach allows a much more accurate modelling of cosmological data, including in particular the complicated systematics they experience.
\item Second, for MCMC methods, the number of data model evaluations is not fixed \textit{a priori}, as some proposed samples are rejected during runtime. One has to assess the convergence of the chain. In contrast, our method only requires a fixed number of realisations to characterise the effective likelihood with the linearised black-box, all of which are used to obtain the inference result. For $S$ target parameters and $N_0$ nuisance parameters realisations at the expansion point, $N_s$ nuisance parameters realisations along each direction in parameter space, one has to perform $N_0 + N_s \times S$ data model evaluations. Increasing $N_0$ or $N_s$ to get a better estimate of the required covariance matrix or gradient only increases the overall computational cost linearly.
\item Third, while MCMC have to be computed sequentially, all the simulations required in the proposed method can be run simultaneously in parallel, or even on different machines. This allows a fast application of the method and makes it particularly suitable for grid computing.
\item Finally, the linearised black-box is trained once and for all independently of the data. This means that if one acquires new data from the same survey, no additional black-box evaluations are required to perform inference, whereas likelihood-based techniques would require a new MCMC. Furthermore, if the cosmological simulations used are stored, they can even be used to perform inference from a different survey, by just replacing the part of the black-box corresponding to survey specifications.
\end{itemize}
{\selfi} also differs from other approaches to ABC (such as likelihood-free rejection sampling, Population Monte Carlo, \textsc{delfi}, or \textsc{bolfi}), which are limited by their inability to scale with the number of target parameters. By relying on an expansion of the simulator, {\selfi} allows the likelihood-free inference of $S \gtrsim 100$ parameters, as is necessary for a model-independent parametrisation of theory in cosmology.

In this work, we demonstrated that a ``non-wiggly'' expansion point $\boldsymbol{\uptheta}_0$ is sufficient to recover the target wiggle function $\boldsymbol{\uptheta}$ in the domain allowed by Planck priors. However, it shall be noted that if the solution is farther from the expansion point, then the method can be iterated. In this case, the posterior mean $\boldsymbol{\upgamma}$ would be used as the new expansion point to train a new linearised black-box, used to obtain a new posterior. Using a sufficient number of iterations, we expect the effective posterior to converge to the true function, even if it strongly deviates from the first expansion point. We leave the detailed investigation of this idea to future studies.

In this paper, we showed a successful application of finite differencing to obtain the gradient of the averaged black-box $\nabla \textbf{f}_0$, but our equations could be used with other techniques, such as automatic differentiation. The data covariance matrix at the expansion point $\textbf{C}_0$ also needs to be evaluated; for this task, and for certain summary statistics, variance reduction techniques such as the use of fixed and paired simulations \citep{AnguloPontzen2016,Villaescusa-Navarro2018} or hybrid estimators \citep{HallTaylor2019} could further be exploited.

As a proof of concept, we applied our technique in conjunction with the artificial galaxy survey simulator \textsc{Simbelmyn\"e}, emulating relevant effects at play: non-linear gravitational structure formation, redshift-space distortions, a survey mask and selection function, and instrumental noise. As a result, the inferred primordial power spectrum is unbiased with a distinct identification of BAO wiggles, even far in the Silk damping tail. We also demonstrated that unbiased inference of cosmological parameters is possible. In spite of the non-linear evolution of structures on small scales, we are able to use the power spectrum of the galaxy field as summary statistic, up to $k_\mathrm{max} = 0.5$~$h$/Mpc. This represents an increase by a factor of $\sim 5$ in the number of modes used with respect to state-of-the-art perturbation theory and backward-modelling techniques, with perspectives for further improvements. Assuming that posteriors are Gaussian and modes are independent, this increase translates into a reduction of the size of credible contours by a factor $\sim \sqrt{5}$.

The data model used in this work remains simplified with respect to some of the complications found in real galaxy surveys. However, due to the flexible nature of the method, it is straightforward to include additional aspects in the inference process: one only has to exchange the black-box for a more sophisticated one. We developed a python code reflecting this versatility, {\pyselfi}, which we publicly released, together with documentation and the data necessary to reproduce the results of the present paper.\footnote{Currently, the code's homepage is hosted at \href{http://pyselfi.florent-leclercq.eu/}{http://pyselfi.florent-leclercq.eu}; the sources are available on GitHub at \href{https://github.com/florent-leclercq/pyselfi/}{https://github.com/florent-leclercq/pyselfi}; and the documentation is on Read the Docs at \href{https://pyselfi.readthedocs.io/en/latest/}{https://pyselfi.readthedocs.io}.} The application of this method to more complex models and to real survey data is left for future research.

In conclusion, the method developed constitutes a computationally efficient and easily applicable framework to infer the primordial matter power spectrum and cosmological parameters from complex black-box mock observations. It allows the use of fully non-linear data models, as required for an optimal analysis of galaxy surveys. Other applications may include the cosmic microwave background, weak gravitational lensing, or the 21~cm signal of hydrogen. The prize for using full forward-modelling in these problems is a potentially vast gain of precision in cosmological constraints.

\appendix
\section{Derivation of the effective likelihood}
\label{apx:Derivation of the effective likelihood}

In this appendix, we derive the approximate likelihood given in equation \eqref{eq:effective_likelihood}, starting from equation \eqref{eq:marginalisation} and the assumptions detailed in section \ref{sssec:The effective likelihood}. We have
\begin{equation}
\p(\boldsymbol{\Phi}, \lbrace \boldsymbol{\Phi}_{\boldsymbol{\uptheta}}^{(i)} \rbrace | \boldsymbol{\uptheta}) \propto \int \p(\boldsymbol{\Phi}|\textbf{s}) \left[\prod_{n=1}^N \p(\boldsymbol{\Phi}_{\boldsymbol{\uptheta}}^{(i)}| \textbf{s})\right] \mathrm{d} \textbf{s}.
\end{equation}
Using the parametric form for $\p(\boldsymbol{\Phi}|\textbf{s})$ (equation \eqref{eq:virtual_signal_pdf}) yields $\p(\boldsymbol{\Phi}, \lbrace \boldsymbol{\Phi}_{\boldsymbol{\uptheta}}^{(i)} \rbrace | \boldsymbol{\uptheta}) \propto \int \exp\left[ \hat{\ell}_1(\boldsymbol{\Phi}, \lbrace \boldsymbol{\Phi}_{\boldsymbol{\uptheta}}^{(i)} \rbrace, \textbf{s}) \right] \, \mathrm{d}\textbf{s}$, with
\begin{eqnarray}
-2\hat{\ell}_1(\boldsymbol{\Phi}, \lbrace \boldsymbol{\Phi}_{\boldsymbol{\uptheta}}^{(i)} \rbrace, \textbf{s}) & \equiv & (\boldsymbol{\Phi}-\textbf{s})^\intercal \boldsymbol{\Sigma}_{\boldsymbol{\uptheta}}^{-1}(\boldsymbol{\Phi}-\textbf{s}) \\
& & + \sum_{i=1}^N (\boldsymbol{\Phi}_{\boldsymbol{\uptheta}}^{(i)}-\textbf{s})^\intercal \boldsymbol{\Sigma}_{\boldsymbol{\uptheta}}^{-1}(\boldsymbol{\Phi}_{\boldsymbol{\uptheta}}^{(i)}-\textbf{s}) \nonumber\\
& & +~(N+1) \log |2\pi \boldsymbol{\Sigma}_{\boldsymbol{\uptheta}}| .\nonumber
\end{eqnarray}
In order to evaluate the integral, we complete the square with respect to $\textbf{s}$ in the argument of the exponential,
\begin{eqnarray}
-2\hat{\ell}_1(\boldsymbol{\Phi}, \lbrace \boldsymbol{\Phi}_{\boldsymbol{\uptheta}}^{(i)} \rbrace, \textbf{s}) & = & \boldsymbol{\Phi}^\intercal \boldsymbol{\Sigma}_{\boldsymbol{\uptheta}}^{-1} \boldsymbol{\Phi} + \sum_{i=1}^N \boldsymbol{\Phi}_{\boldsymbol{\uptheta}}^{(i)\intercal} \boldsymbol{\Sigma}_{\boldsymbol{\uptheta}}^{-1} \boldsymbol{\Phi}_{\boldsymbol{\uptheta}}^{(i)} \nonumber\\
& & -~2(\boldsymbol{\Phi} + N\boldsymbol{\hat{\Phi}}_{\boldsymbol{\uptheta}})^\intercal \boldsymbol{\Sigma}_{\boldsymbol{\uptheta}}^{-1} \textbf{s} \nonumber\\
& & +~(N+1) \, \textbf{s}^\intercal \boldsymbol{\Sigma}_{\boldsymbol{\uptheta}}^{-1} \textbf{s} \nonumber\\
& & +~(N+1) \log |2\pi \boldsymbol{\Sigma}_{\boldsymbol{\uptheta}}|  \\
& = & \boldsymbol{\Phi}^\intercal \boldsymbol{\Sigma}_{\boldsymbol{\uptheta}}^{-1} \boldsymbol{\Phi} + \sum_{i=1}^N \boldsymbol{\Phi}_{\boldsymbol{\uptheta}}^{(i)\intercal} \boldsymbol{\Sigma}_{\boldsymbol{\uptheta}}^{-1} \boldsymbol{\Phi}_{\boldsymbol{\uptheta}}^{(i)} \nonumber\\
& & -~\boldsymbol{\upeta}^\intercal (N+1) \, \boldsymbol{\Sigma}_{\boldsymbol{\uptheta}}^{-1} \boldsymbol{\upeta} \nonumber\\
& & +~(N+1)\bigl[(\textbf{s} - \boldsymbol{\upeta})^\intercal \, \boldsymbol{\Sigma}_{\boldsymbol{\uptheta}}^{-1} (\textbf{s}-\boldsymbol{\upeta}) \nonumber\\
& & \quad +~  \log |2\pi \boldsymbol{\Sigma}_{\boldsymbol{\uptheta}}|\bigr], \nonumber
\end{eqnarray}
where we have recognised $\boldsymbol{\hat{\Phi}}_{\boldsymbol{\uptheta}} = \frac{1}{N} \sum_{i=1}^N \boldsymbol{\Phi}_{\boldsymbol{\uptheta}}^{(i)}$ (equation \eqref{eq:estimated_mean}) and introduced $\boldsymbol{\upeta} \equiv (\boldsymbol{\Phi} + N\boldsymbol{\hat{\Phi}}_{\boldsymbol{\uptheta}})/(N+1)$. After integration over $\textbf{s}$, the last term gives a constant factor, so that $\p(\boldsymbol{\Phi}, \lbrace \boldsymbol{\Phi}_{\boldsymbol{\uptheta}}^{(i)} \rbrace | \boldsymbol{\uptheta}) \propto \exp\left[ \hat{\ell}_2(\boldsymbol{\Phi}, \lbrace \boldsymbol{\Phi}_{\boldsymbol{\uptheta}}^{(i)} \rbrace) \right]$, with
\begin{eqnarray}
-2\hat{\ell}_2(\boldsymbol{\Phi}, \lbrace \boldsymbol{\Phi}_{\boldsymbol{\uptheta}}^{(i)} \rbrace) & \equiv & \boldsymbol{\Phi}^\intercal \boldsymbol{\Sigma}_{\boldsymbol{\uptheta}}^{-1} \boldsymbol{\Phi} + \sum\limits_{i=1}^N \boldsymbol{\Phi}_{\boldsymbol{\uptheta}}^{(i)\intercal} \boldsymbol{\Sigma}_{\boldsymbol{\uptheta}}^{-1} \boldsymbol{\Phi}_{\boldsymbol{\uptheta}}^{(i)} \\
& & -(\boldsymbol{\Phi}+N\boldsymbol{\hat{\Phi}}_{\boldsymbol{\uptheta}})^\intercal \frac{1}{N+1} \, \boldsymbol{\Sigma}_{\boldsymbol{\uptheta}}^{-1} (\boldsymbol{\Phi}+N\boldsymbol{\hat{\Phi}}_{\boldsymbol{\uptheta}}) . \nonumber
\end{eqnarray}
We now complete the square with respect to $\boldsymbol{\Phi}$ to obtain
\begin{eqnarray}
-2\hat{\ell}_2(\boldsymbol{\Phi}, \lbrace \boldsymbol{\Phi}_{\boldsymbol{\uptheta}}^{(i)} \rbrace) & = & (\boldsymbol{\Phi} - \boldsymbol{\hat{\Phi}}_{\boldsymbol{\uptheta}})^\intercal \left( \frac{N+1}{N} \boldsymbol{\Sigma}_{\boldsymbol{\uptheta}} \right)^{-1}  (\boldsymbol{\Phi} - \boldsymbol{\hat{\Phi}}_{\boldsymbol{\uptheta}}) \nonumber\\
& & +~\mathrm{constant~terms.}
\label{eq:effective_likelihood_intermediate}
\end{eqnarray}
In order to obtain a computable approximation of the likelihood, the unknown covariance $\boldsymbol{\Sigma}_{\boldsymbol{\uptheta}}$ in $\hat{\ell}_2(\boldsymbol{\Phi}, \lbrace \boldsymbol{\Phi}_{\boldsymbol{\uptheta}}^{(i)} \rbrace)$ has to be approximated by $\boldsymbol{\hat{\Sigma}}_{\boldsymbol{\uptheta}}$, defined by equation \eqref{eq:estimated_covariance}. The covariance of the effective likelihood is therefore $\boldsymbol{\hat{\Sigma}}_{\boldsymbol{\uptheta}}' \equiv \frac{N+1}{N}\boldsymbol{\hat{\Sigma}}_{\boldsymbol{\uptheta}}$. The unknown inverse covariance $\boldsymbol{\Sigma}_{\boldsymbol{\uptheta}}^{-1}$ is also replaced by its unbiased computable approximation $\boldsymbol{\hat{\Sigma}}_{\boldsymbol{\uptheta}}^{-1}$, defined by equation \eqref{eq:estimated_inverse_covariance}. Finally, we use the normalisation condition $\int \p(\boldsymbol{\Phi} | \lbrace \boldsymbol{\Phi}_{\boldsymbol{\uptheta}}^{(i)} \rbrace , \boldsymbol{\uptheta}) \, \mathrm{d}\boldsymbol{\Phi} = \int \frac{\p(\boldsymbol{\Phi} , \lbrace \boldsymbol{\Phi}_{\boldsymbol{\uptheta}}^{(i)} \rbrace | \boldsymbol{\uptheta})}{\p(\lbrace \boldsymbol{\Phi}_{\boldsymbol{\uptheta}}^{(i)} \rbrace | \boldsymbol{\uptheta})} \, \mathrm{d}\boldsymbol{\Phi} = 1$ and evaluate at $\boldsymbol{\Phi}=\boldsymbol{\Phi}_\mathrm{O}$, as prescribed by equation \eqref{eq:effective_likelihood_def}, to obtain $\hat{\ell}^N(\boldsymbol{\uptheta})$ given by equation \eqref{eq:effective_likelihood}. When $N \rightarrow \infty$, $\boldsymbol{\hat{\Phi}}_{\boldsymbol{\uptheta}} \longrightarrow \textbf{s}$ and $\boldsymbol{\hat{\Sigma}}_{\boldsymbol{\uptheta}}' \longrightarrow \boldsymbol{\Sigma}_{\boldsymbol{\uptheta}}$, thus the limiting approximation is $\tilde{\ell}(\boldsymbol{\uptheta})$ given by equation \eqref{eq:effective_likelihood_limiting}.

\section{Derivation of the effective posterior}
\label{apx:Derivation of the effective posterior}

We recall the canonical form of the Gaussian distribution with mean $\textbf{x}_0$ and covariance matrix $\textbf{X}$, given as
\begin{eqnarray}
-2\log \p(\textbf{x}) & = & \log\left| 2\pi\textbf{X} \right| + (\textbf{x}-\textbf{x}_0)^\intercal \textbf{X}^{-1} (\textbf{x}-\textbf{x}_0) \\
& = & \log\left| 2\pi\textbf{X} \right| + \boldsymbol{\upxi}_0^\intercal \textbf{X} \boldsymbol{\upxi}_0 -2 \boldsymbol{\upxi}_0^\intercal \textbf{x} + \textbf{x}^\intercal \textbf{X}^{-1} \textbf{x}, \nonumber
\end{eqnarray}
where $\boldsymbol{\upxi}_0 \equiv \textbf{X}^{-1} \textbf{x}_0$.

Using the linearised data model (equation \eqref{eq:linearised_black_box}) in the expression of the effective likelihood (equation \eqref{eq:linearised_effective_likelihood}), we get
\begin{eqnarray}
-2\hat{\ell}^N(\boldsymbol{\uptheta}) & = & \log \left| 2\pi \textbf{C}_0 \right| + \left[ \boldsymbol{\Phi}_\mathrm{O} - \textbf{f}_0 - \nabla \textbf{f}_0 \cdot (\boldsymbol{\uptheta}-\boldsymbol{\uptheta}_0 )\right]^\intercal \cdot \nonumber\\
& & \quad\quad\quad\quad \textbf{C}_0^{-1} \left[ \boldsymbol{\Phi}_\mathrm{O} - \textbf{f}_0 - \nabla \textbf{f}_0 \cdot (\boldsymbol{\uptheta}-\boldsymbol{\uptheta}_0 )\right] \nonumber\\
& = & \log \left| 2\pi \textbf{C}_0 \right| + (\textbf{y}_0 - \boldsymbol{\uptheta})^\intercal \textbf{N}_0^{-1} (\textbf{y}_0 - \boldsymbol{\uptheta}),
\end{eqnarray}
where we have defined
\begin{equation}
\textbf{N}_0 \equiv \left[ (\nabla \textbf{f}_0)^\intercal \textbf{C}_0^{-1} \nabla \textbf{f}_0 \right]^{-1}
\end{equation}
and
\begin{equation}
\textbf{y}_0 \equiv \boldsymbol{\uptheta}_0 + (\nabla \textbf{f}_0)^{-1} \cdot (\boldsymbol{\Phi}_\mathrm{O}-\textbf{f}_0),
\end{equation}
$(\nabla \textbf{f}_0)^{-1}$ denoting the adjoint of the Jacobian characterising the linearised black-box (its computation will not be necessary). In canonical form, the Gaussian effective likelihood is written
\begin{equation}
-2\hat{\ell}^N(\boldsymbol{\uptheta}) = \log \left| 2\pi \textbf{C}_0 \right| + \boldsymbol{\upmu}_0^\intercal \textbf{N}_0 \boldsymbol{\upmu}_0 - 2\boldsymbol{\upmu}_0 \boldsymbol{\uptheta} + \boldsymbol{\uptheta}^\intercal \textbf{N}_0^{-1} \boldsymbol{\uptheta},
\end{equation}
with $\boldsymbol{\upmu}_0 \equiv \textbf{N}_0^{-1} \textbf{y}_0 = \textbf{N}_0^{-1}\boldsymbol{\uptheta}_0 + (\nabla \textbf{f}_0)^\intercal \textbf{C}_0^{-1} (\boldsymbol{\Phi}_\mathrm{O}-\textbf{f}_0)$.
Similarly, the prior (equation \eqref{eq:prior}) is written
\begin{equation}
-2\log \p(\boldsymbol{\uptheta}) = \log\left| 2\pi\textbf{S} \right| + \boldsymbol{\upeta}_0^\intercal \textbf{S} \boldsymbol{\upeta}_0 -2 \boldsymbol{\upeta}_0^\intercal \boldsymbol{\uptheta} + \boldsymbol{\uptheta}^\intercal \textbf{S}^{-1} \boldsymbol{\uptheta}
\end{equation}
where $\boldsymbol{\upeta}_0 \equiv \textbf{S}^{-1} \boldsymbol{\uptheta}_0$.

Adding the two expressions, we find that the effective posterior verifies
\begin{eqnarray}
-2\log \p(\boldsymbol{\uptheta}|\boldsymbol{\Phi})_{|\boldsymbol{\Phi}=\boldsymbol{\Phi}_\mathrm{O}} & = & -2(\boldsymbol{\upmu}_0+\boldsymbol{\upeta}_0)^\intercal \boldsymbol{\uptheta} \nonumber\\
& & + \boldsymbol{\uptheta}^\intercal (\textbf{N}_0^{-1} + \textbf{S}^{-1}) \boldsymbol{\uptheta} \nonumber\\
& & +~\mathrm{constant~terms.}
\end{eqnarray}
This is the canonical form of a Gaussian distribution, where the covariance matrix is identified as $\boldsymbol{\Gamma} \equiv (\textbf{N}_0^{-1} + \textbf{S}^{-1})^{-1}$, giving equation \eqref{eq:filter_var}, and the mean is identified as
\begin{eqnarray}
\boldsymbol{\upgamma} & = & \boldsymbol{\Gamma}(\boldsymbol{\upmu}_0+\boldsymbol{\upeta}_0) \\
& = & \boldsymbol{\Gamma}\textbf{N}_0^{-1}\boldsymbol{\uptheta}_0 + \boldsymbol{\Gamma}(\nabla \textbf{f}_0)^\intercal \textbf{C}_0^{-1} (\boldsymbol{\Phi}_\mathrm{O}-\textbf{f}_0) + \boldsymbol{\Gamma}\textbf{S}^{-1}\boldsymbol{\uptheta}_0, \nonumber
\end{eqnarray}
giving equation \eqref{eq:filter_mean}.

Note that the above calculation is analogous to the derivation of the Wiener filter equations: assuming a linear data model ($\textbf{d}=\textbf{s}+\textbf{n}$), a prior with mean $\bar{\textbf{s}}$ and signal covariance $\textbf{S}$, and a likelihood with mean $\bar{\textbf{d}}$ and noise covariance $\textbf{N}$, the filter covariance is $(\textbf{N}^{-1}+\textbf{S}^{-1})^{-1}$ and the filtered signal is $\bar{\textbf{s}} + (\textbf{N}^{-1}+\textbf{S}^{-1})^{-1}\textbf{N}^{-1}(\textbf{d}_\mathrm{O}-\bar{\textbf{d}})$.

\section*{Statement of contribution}

Study concept and design (JJ, FL, WE); design of prior optimisation and cosmological parameter inference (FL); original code implementation of the filter equations (WE); code rewriting and enhancements (FL); design and implementation of the data model (FL); running of the simulations (FL); drafting of the manuscript (WE); critical revision of the manuscript (FL); proofreading (FL, WE, JJ, AH); supervision (JJ, FL); support and interpretation of results (JJ, AH). All authors read and approved the final manuscript.

\acknowledgments
FL is grateful to Guilhem Lavaux and Andrew Jaffe for useful discussions. This work has made use of a modified version of \href{https://pypi.org/project/pyGTC/}{\textsc{pygtc}} \citep{Bocquet2016}. Numerical computations were done on the cx1 cluster hosted by the Research Computing Service facilities at Imperial College London (\href{http://doi.org/10.14469/hpc/2232}{doi:10.14469/hpc/2232}). This work is done within the Aquila Consortium (\href{https://aquila-consortium.org}{https://aquila-consortium.org}).

FL acknowledges funding from the Imperial College London Research Fellowship Scheme. This research was supported by the DFG cluster of excellence ``Origin and Structure of the Universe'' (\href{www.universe-cluster.de}{www.universe-cluster.de}).

\renewcommand{\emph}[1]{\textit{#1}}

\section*{References}
\bibliography{/home/leclercq/workspace/biblio/biblio}

\end{document}